\documentclass[10pt]{article}

 \usepackage{amsmath,amsthm,amssymb}
 \usepackage{makeidx,epsfig,lscape}
 \usepackage{color,colortbl}
 \usepackage{fancyhdr}
 \usepackage{enumerate}
 \usepackage{xcolor,pict2e}
\usepackage{amsmath}
 \definecolor{myaqua}{rgb}{0.0,0.5,0.55}
 \definecolor{lightaqua}{rgb}{0.75,0.95,0.95}

 \usepackage[colorlinks = true,
            linkcolor = myaqua,
            urlcolor  = black,
            citecolor = myaqua]{hyperref}

\usepackage{floatrow}
\def\bt{\begin{tabular}}
\def\et{\end{tabular}}
\def\and{\mbox{ and }}

\def\1{{\bf 1}}

\begin{document}
\begin{center}
{ 


{\noindent{\Large\bf
   Poisson-modification of the Quasi Lindley distribution and its zero modification for over-dispersed count data}}
  \\[6mm]
{ Ramajeyam Tharshan$^{1,2}$ $\&$ Pushpakanthie Wijekoon$^{3}$}}
\\[2mm]
\
{ 
$^1$Postgraduate Institute of Science, University of Peradeniya, Peradeniya, Sri Lanka\\[1mm]
$^2$Department of Mathematics and Statistics, University of Jaffna, Jaffna, Sri Lanka\\[1mm]
$^3$Department of Statistics and Computer Science, University of Peradeniya, Peradeniya,Sri Lanka}\\[1mm]
\end{center}
{\noindent{\large \bf Abstract}{\\[3mm]
 \textup{
In this paper, an alternative mixed Poisson distribution is proposed by amalgamating Poisson distribution and a modification of the Quasi Lindley distribution. Some fundamental structural properties of the new distribution, namely the shape of the distribution and moments and related measures, are explored. It was noted that the new distribution to be either unimodal or bimodal, and over-dispersed. Further, it has a tendency to accommodate various right tail behaviors and variance-to-mean ratios. A zero-modified version of this distribution is also derived. Its unknown parameter estimation by using the maximum likelihood estimation method is examined by a simulation study based on the asymptotic theory. Finally, three real-world data sets are used to illustrate the flexibility and potentiality of the new distribution. 
 }}}
 \\[4mm]
 {\noindent{\large\bf Keywords}{ \\[3mm]
 over-dispersion, mixed Poisson distribution, zero modification, Lindley distribution, goodness of fit.}
\section{Introduction}
The Poisson distribution is a standard tool to model the count data if the empirical and theoretical properties satisfy the related underline assumptions.  A random variable $X$ is said to have a Poisson distribution with parameter $\lambda$ if both the $E(X)$ and $Var(X)$ of the distribution equal to the parameter $\lambda$. This property is commonly known as equidispersion.
Even though its probability mass function (pmf) is very flexible to compute its probabilities, in some real-world applications, especially, actuarial, ecology, and genetics, the Poisson distribution fails to match empirical observations. Here the variance of the observed data exceeds the theoretical variance. This phenomenon is explained as over-dispersion or variation inflation (Greenwood and Yule, 1920). The over-dispersion occurs by the failure of the basic assumptions of the Poisson distribution. The reasons might be by phenomena of the clustered structure of the population or population is heterogeneous, and heavy right tail that cannot accommodate by the Poisson distribution (McCullagh et al., 1989; Ridout et al., 1998). The heterogeneity of a population is determined by the Poisson parameter $\lambda$ which differs individual to individual, and then the $Var(X)=\tau E(X); (\tau>1)$, where $\tau$ is called the index of dispersion.\\\\ The mixed Poisson distributions are well-known flexible modeling methods to explain the heterogeneity of the Poisson parameter as well as heavy right tail behaviors (Feller,1943; Shaked, 1980). The mixed Poisson distribution is a resultant distribution or unconditional distribution by assuming that the Poisson parameter is a random variable that has as a parameterized distribution $P$. The distribution $P$ and its parameter vector $\Theta$ are called prior distribution and hyperparameter, respectively. Then, the resultant distribution of the random variable $X$ can be expressed mathematically in the following form\\
 \begin{equation}
     f_{X}(x)=\int_{0}^{\infty}f_{X|\Lambda}(x|\lambda)f_{\Lambda}^{P}(\lambda)d\lambda,
 \end{equation}
 where  $X|\Lambda$ has a Poisson distribution with parameter $\lambda$
 as
 \begin{equation}
f_{X|\Lambda}(x|\lambda)=\displaystyle\frac{e^{-\lambda}\lambda^{x}}{x!} ,\hspace{0.1cm} x=0,1,2,...,\hspace{0.1cm} \lambda>0,
\end{equation} 
 $\Lambda$ is the random variable of the Poisson parameter $\lambda$, and $f_{\Lambda}^{P}(\lambda)$ is the density function of the assumed continuous distribution $P$  to the Poisson parameter $\lambda$. Hence, the random variable
 $X$ has the same support of $X|\Lambda$ with parameter(s) of the prior distribution. Further, Lynch (1988) showed that the form of the mixing distribution has ascendancy over to the form of the resultant mixed distribution. \\\\ In literature,  researchers assumed the standard lifetime distributions to model the Poisson parameter $\lambda$ as a classical approach. Greenwood and Yule (1920) used the gamma distribution, and the resultant distribution is negative binomial (NBD). Johnson et al., (1992) assumed the exponential distribution to model the Poisson parameter, and the resultant distribution is geometric distribution (GD). Even though NBD and GD are computationally flexible pmfs, they are not befitting distributions for a higher value of $\tau$ and long right tail. In this context, researchers have assumed several modifications of the standard lifetime distributions for the Poisson parameter. They were modified to have more flexibility in their shapes and failure rate criteria than the standard lifetime distributions. Confluent Hypergeometric series, Gamma product ratio, Generalized gamma, Shifted gamma, Inverse gamma, and Modified Bessel of the 3rd kind are used by Bhattacharya  (1967), Irwin (1975), Albrecht (1984), Ruohonen (1988), Willmot (1993), and Ong and Muthaloo (1995), respectively to model the Poisson parameter. The pmfs of such distributions are derived through the recursive formulas or Laplace transform technique, or by using the special mathematical functions. Hence, computing the probabilities of such distributions is complicated and they are limited in practice. \\\\The Lindley distribution (LD) is one of the finite mixture distributions introduced by Lindley (1958) having the density function
\begin{equation}
    f_{\Lambda}(\lambda)=\frac{\theta^2}{1+\theta}(1+\lambda)e^{-\theta \lambda} \hspace*{0.1cm},\hspace*{0.3cm} \lambda>0,\hspace*{0.1cm} \theta>0,
\end{equation}
where $\theta$ is the shape parameter, and $\Lambda$ is the respective random variable. Equation (3) presents a two-component mixture of two different continuous distributions namely exponential $(\theta)$ and gamma $(2,\theta)$ distributions with the mixing proportion, $p=\displaystyle\frac{\theta}{1+\theta}$. Sankaran (1970) introduced the one-parameter discrete Poisson-Lindley distribution (PLD) by combining the Poisson and LD. Its pmf is given as
\begin{equation}
    f_{X}(x)=\frac{\theta^2(x+\theta+2)}{(\theta+1)^{x+3}},\hspace*{0.1cm} x=0,1,2,..., \hspace*{0.1cm}\theta>0.
\end{equation}
Note that the pmf of the PLD is an explicit form. Then, obtaining its probabilities is computationally flexible. Further, there are some extensions of the PLD to account for the additional heterogeneity by the zero inflation/deflation.  Ghitany et al. (2008) proposed the zero-truncated PLD and Xavier et al. (2018) introduced the zero-modified PLD.  However, the PLD flexibility is limited to fit various types of the over-dispersed count data sets since it has only one parameter. Then, as an alternative to PLD, Bhati et al. (2015) have obtained the Generalized Poisson-Lindley distribution (GPLD), where the Poisson parameter is distributed to Two-parameter Lindley distribution (Shanker et al., 2013b); Wongrin and Bodhisuwan (2016) introduced Poisson-generalized Lindley distribution (PGLD), which was obtained by mixing the Poisson distribution
with the generalized Lindley distribution (Elbatal et al., 2013); Grine et al. (2017) have obtained the Quasi-Poisson distribution (PQLD) by modeling Poisson parameter to Quasi Lindley distribution (Shanker et al., 2013a). Table 1 summarizes the mixing proportions, mixing components, and parameters of the above continuous distributions that have been used to model the Poisson parameter. We can see that the mixing proportions of the Two-parameter Lindley and generalized Lindley distributions are incorporated with the scale parameter, $\theta$ of the mixing components. Further, the shape parameter of the mixing component gamma $(2,\theta)$ is fixed with value 2 for the Two-parameter Lindley and the Quasi Lindley distributions. These settings of such mixing distributions may limit the flexibility of the above-mentioned Poisson mixtures to fit well for the various types of the right tail heaviness and $\tau$ for an over-dispersed count data (Tharshan and Wijekoon, 2020 a b).\\\\\\\\
Table 1. Mixing proportions, mixing components, and parameters of some modified-Lindley distributions. 
\begin{table}[h!]
\centering
\begin{small}
  \begin{tabular}{ ccccc} 
  \hline
  Distribution&Mixing proportion&Mixing components&\multicolumn{2}{c}{Parameters} \\
  \hline
  &&&shape&scale\\
  \hline
    &&&&\\
  Two-parameter Lindley&$\displaystyle\frac{\theta}{\theta+\alpha}$&exponential $(\theta)$ , gamma $(2, \theta)$&$\theta, \alpha$&\\
    &&&&\\
  Generalized Lindley&$\displaystyle\frac{\theta}{1+\theta}$&gamma $(\alpha,\theta)$ , gamma $(\beta, \theta)$&$\theta, \alpha,\beta$&\\
    &&&&\\
  Quasi Lindley&$\displaystyle\frac{\alpha}{\alpha+1}$&exponential $(\theta)$ , gamma $(2, \theta)$&$\theta$&$\alpha$\\
    &&&&\\
  \hline
\end{tabular}
\end{small}
\end{table}
\\
The main contribution of this paper is to propose an alternative mixed Poisson distribution and its zero-modified version for over-dispersed count data to address the above issues. It is obtained by mixing the Poisson distribution with the Modification of the Quasi Lindley distribution (MQLD) (Tharshan and Wijekoon, 2021). The density function of the MQLD$(\theta,\alpha,\delta)$ is given as
\begin{equation}
    f_{\Lambda}(\lambda; \theta,\alpha,\delta)=\displaystyle\frac{\theta e^{-\theta \lambda}}{(\alpha^3+1)\Gamma(\delta)}\bigg(\Gamma(\delta)\alpha^3 + (\theta \lambda)^{\delta-1}\bigg) ;\lambda>0, \theta>0, \alpha^3>-1, \delta>0,
\end{equation}
where $\alpha$, and $\delta$ are shape parameters, $\theta$ is a scale parameter, and $\Lambda$ is the respective random variable. Equation (5) presents the mixture of two non-identical distributions, exponential $(\theta)$ , and gamma $(\delta, \theta)$ with the mixing proportion, $p=\displaystyle\frac{\alpha^3}{\alpha^3+1}$. We can clearly observe that its mixing proportion $p$ does not incorporate with scale parameter, $\theta$ of the mixing components. Further, the shape parameter of the mixing component gamma distribution, $\delta$ is not fixed with a value. The authors have shown that these settings of the MQLD provide the capability to capture the various ranges of right tail heaviness measured by excess kurtosis (kurtosis -3), horizontal symmetry measured by skewness, and heterogeneity measured by Fano factor (variance-to-mean ratio) by setting its parameter values. Further, its density function can be either unimodal or bimodal.\\\\
 The remaining part of this paper is organized as follows: In section 2, we introduce the PMQLD with its explicit forms of probability mass and distribution functions. Its fundamental structural properties are discussed in section 3. Section 4 presents the zero-modified version of PMQLD. The simulation of its random variables and parameter estimations are discussed in section 5. Finally, a simulation study is done to examine the performance of parameter estimation by using the maximum likelihood estimation method, and some real-world examples are taken to show the applicability of the proposed model by comparing it with some other existing Poisson mixtures, NBD, GD, PLD, GPLD, PQLD, PGLD. 
 \section{Formulation of the new mixed Poisson distribution}
 In this section, we introduce the new mixed Poisson distribution with its pmf and cumulative distribution function (cdf). \\\\Let the random variable $X$ represent the total counts of a specific experiment with mean $\lambda$. Then, the traditional distribution to calculate probabilities of such outcomes is the Poisson distribution. The PMQLD is obtained by mixing the Poisson with MQLD (Tharshan and Wijekoon, 2021) for over-dispersed count data.  The following proposition gives the pmf of the PMQLD.\\\\\textbf{Proposition 1.} Let $X|\Lambda$ is a random variable that follow the Poisson distribution with parameter $\lambda$, abbreviated as $X|\Lambda \sim$ Poisson $(\lambda)$ and the Poisson parameter $\Lambda \sim $ MQLD $(\theta,\alpha,\delta)$. Then, the pmf of the PMQLD is defined as
\begin{equation}
 f_{X}(x)=\displaystyle\frac{\theta\bigg(\Gamma(\delta)\Gamma(x+1)\alpha^3(1+\theta)^{\delta-1}+\theta^{\delta-1}\Gamma(x+\delta)\bigg)}{x !(\alpha^3+1)  (1+\theta)^{x+\delta}\Gamma(\delta)} , x=0,1,2, . . .,  \theta>0, \delta>0, \alpha^3>-1. 
 \end{equation}
 \textbf{Proof:}\\\\Since $X|\Lambda \sim$ Poisson $(\lambda)$  and $\Lambda \sim $ MQLD $(\theta, \alpha, \delta)$, the unconditional distribution of $X$ can be obtained by substituting equations (2) and (5) in equation (1) as below \begin{center}$f(X)=\displaystyle\int_{0}^{\infty} \displaystyle\frac{e^{-\lambda}\lambda^{x}}{x!}\displaystyle\frac{\theta e^{-\theta \lambda}}{(\alpha^3+1)\Gamma(\delta)}\bigg(\Gamma(\delta)\alpha^3 + (\theta \lambda)^{\delta-1}\bigg)d\lambda$\end{center} \begin{center}$=\displaystyle\frac{\theta}{x!(\alpha^3+1) \Gamma(\delta)}\bigg(\Gamma(\delta)\alpha^3\displaystyle\int_{0}^{\infty} e^{-\lambda(1+\theta)} \lambda^{x}d\lambda+ \theta^{\delta-1}\displaystyle\int_{0}^{\infty}\lambda^{x+\delta-1}e^{-\lambda(1+\theta)}d\lambda\bigg)$\end{center}
 \begin{center}$=\displaystyle\frac{\theta}{x!(\alpha^3+1) \Gamma(\delta)}\bigg(\displaystyle\frac{\Gamma(\delta)\alpha^3\Gamma(x+1)}{(1+\theta)^{x+1}}+\displaystyle\frac{\theta^{\delta-1}\Gamma(x+\delta)}{(1+\theta)^{x+\delta}}\bigg)$\end{center}
 \begin{center}$=\displaystyle\frac{\theta}{x!(\alpha^3+1) \Gamma(\delta)(1+\theta)^{x+\delta}}\bigg((1+\theta)^{\delta-1}\Gamma(\delta)\alpha^3\Gamma(x+1)+\theta^{\delta-1}\Gamma(x+\delta)\bigg).$
\end{center}
\textbf{Remarks:}
\begin{enumerate}
\item Equation (6) presents a two-component mixture of GD$(\frac{\theta}{1+\theta})$ and NBD$(\delta,\frac{1}{1+\theta})$ with the mixing proportion $p=\frac{\alpha^3}{\alpha^3+1}$. 
\item For $\alpha \rightarrow 0$, the PMQLD reduces to the NBD$(\delta,\frac{1}{1+\theta})$. 
\item For $\alpha \rightarrow \infty$, the PMQLD reduces to the GD$(\frac{\theta}{1+\theta})$. 
\end{enumerate}
The right tail behaviors of the PMQLD for different values of $\theta, \alpha$, and $\delta$ are illustrated in Figure 1. For fixed $\alpha$ and $\delta$, it is clear that the distribution's right tail approaches to zero at a faster rate when $\theta$ increases. For fixed $\theta$ and $\delta$, and when $\alpha$ is increasing, the distribution's right tail approaches to zero at a slower rate when compared with the changes of $\theta$. Further, for fixed $\theta$ and $\alpha$, and when $\delta$ is increasing, the distribution captures more right tail. Figure 2 shows the right tail behavior of the PMQLD when the parameters $\theta$ and $\delta$ tend to be very smaller and higher values. From Figure 2 (a), (b), for given values of $\alpha$ and $\delta$, we can observe that the distribution's right tail approaches to zero at a slower rate when $\theta$ is equal to a very smaller value. However, it is at a faster rate when $\theta$ is equal to a very higher value. Further, from Figure 2 (c), (d), it can be noticed that this behavior is reversed for $\delta$ when compared with $\theta$ values.   From Figure 3 , we may note that the PMQLD may be a bimodal distribution when parameter value $\delta$ is very different (higher value) from the parameter values of $\theta$ and $\alpha.$\\\\The corresponding cumulative distribution function of PMQLD is given as
\begin{equation}
    F_{X}(x)=\displaystyle\sum_{t=0}^{x}f(x)=\displaystyle\frac{\delta(1+\theta)^{\delta-1}\Gamma(\delta)\alpha^3\Gamma(x+1)((1+\theta)^{x+1}-1)+\theta^\delta\Gamma(x+\delta+1)_{2}F_{1}(1,x+\delta+1;\delta+1;\frac{\theta}{1+\theta})}{(\alpha^3+1)\Gamma(\delta)x!\delta(1+\theta)^{x+\delta+1}}
\end{equation}\\$\hspace*{10.5cm},x=0,1,2, . . .,  \theta>0, \delta>0, \alpha^3>-1,$\\\\
where $_{2}F_{1}(c,d;r;w)$ is the Gaussian hypergeometric function defined as \begin{center}  $_{2}F_{1}(c,d;r;w)=\displaystyle\sum_{i=0}^{\infty}\frac{(c)_{i}(d)_{i}w^{i}}{(r)_{i}i!},$ \end{center} which is a special case of the generalized hypergeometric function given by the
expression\begin{center} $_{a}F_{b}(p_{1},p_{2},...p_{a};q_{1},q_{2},...q_{b};w)=\displaystyle\sum_{i=0}^{\infty}\frac{(p_{1})_{i}...(p_{a})_{i}w^{i}}{(q_{1})_{i}...(q_{b})_{i}i!},\hspace*{0.3cm} $\end{center} and $(p)_{i}=\frac{\Gamma(p+i)}{\Gamma(p)}=p(p+1)...(p+i+1)$ is the  Pochhammer symbol.
 \begin{figure}[h!]\label{fig:my_label}
        \includegraphics[width=15cm, height=12.5cm]{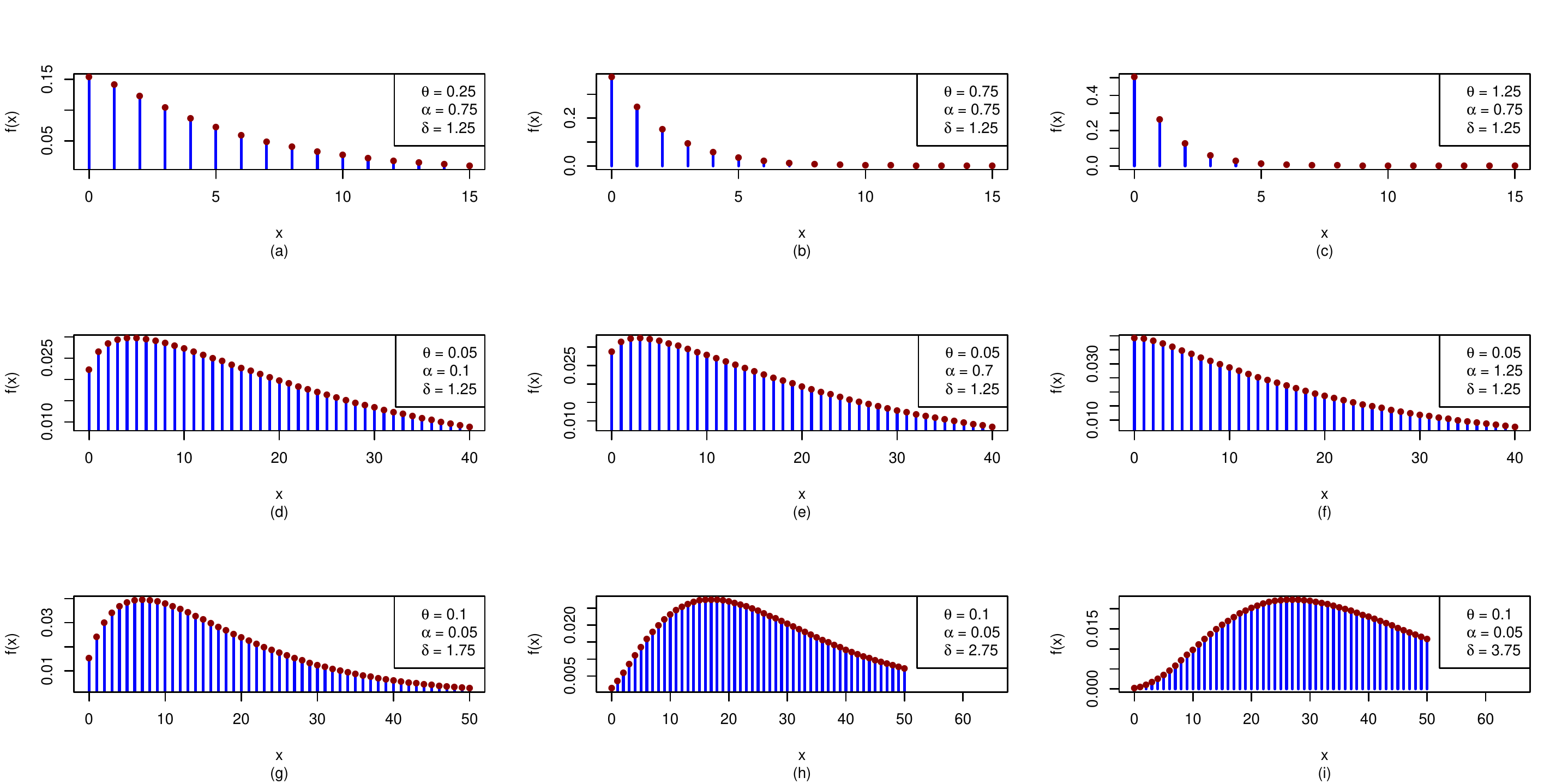}
   \caption{The probability mass function of the PMQLD at different parameter values of $\theta,\alpha,$and ,$\delta$}
   \floatfoot{(a), (b), and (c): $\alpha,$ and $\delta$ are fixed, and $\theta$ values are changed; (d), (e), and (f) : $\theta,$ and $\delta$ are fixed, and $\alpha$ values are changed; (g), (h), and (i): $\theta,$ and $\alpha,$ are fixed, and $\delta$ values are changed.}
   \end{figure}
   \begin{figure}[h!]\label{fig:my_label}
        \includegraphics[width=15.0cm, height=9.0cm]{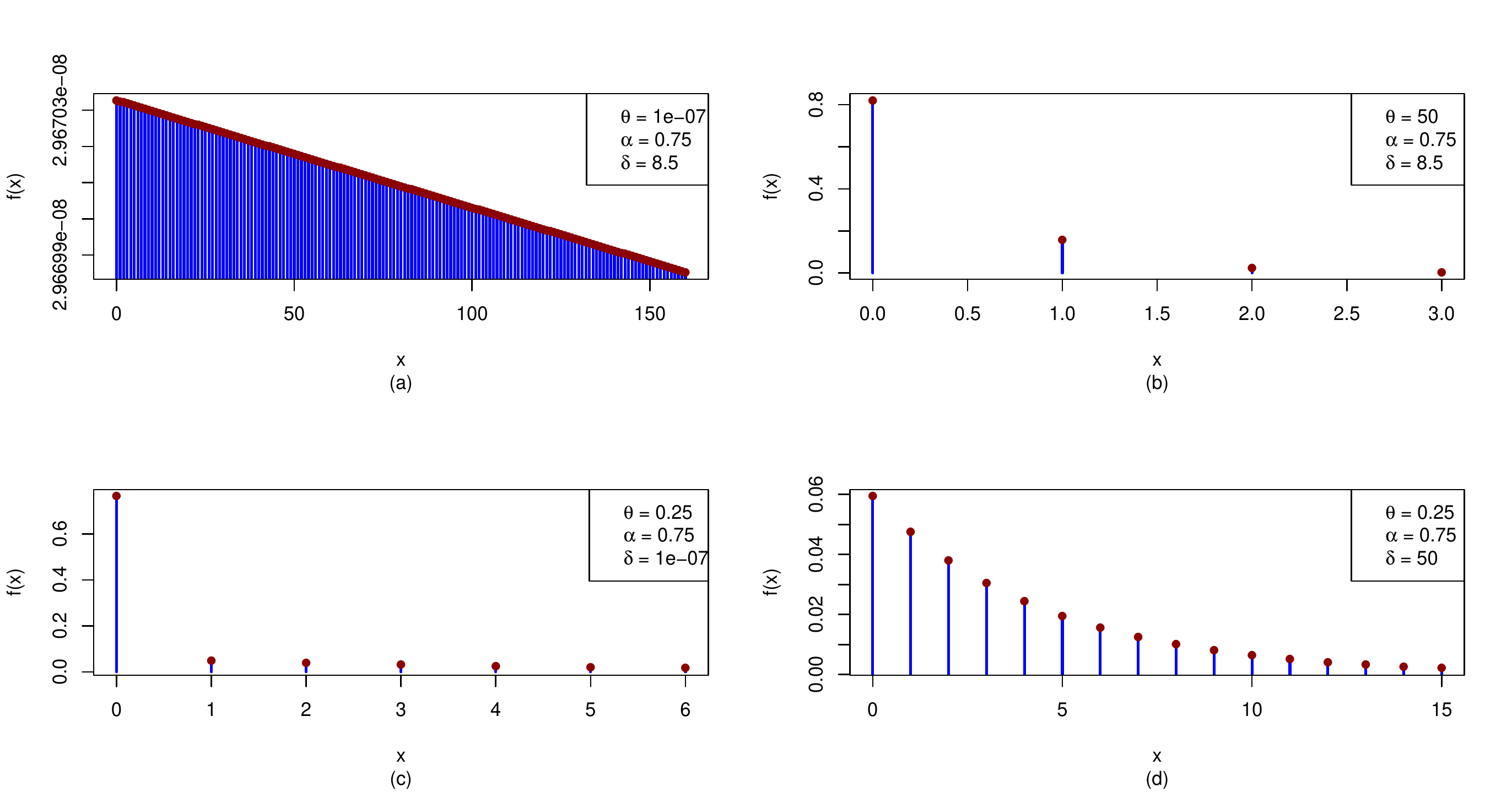}
   \caption{The probability mass function of the PMQLD for very smaller and higher values of $\theta$ and $\delta$}
   \end{figure}
\begin{figure}[h!]\label{fig:my_label}
        \includegraphics[width=15.5cm, height=10.0cm]{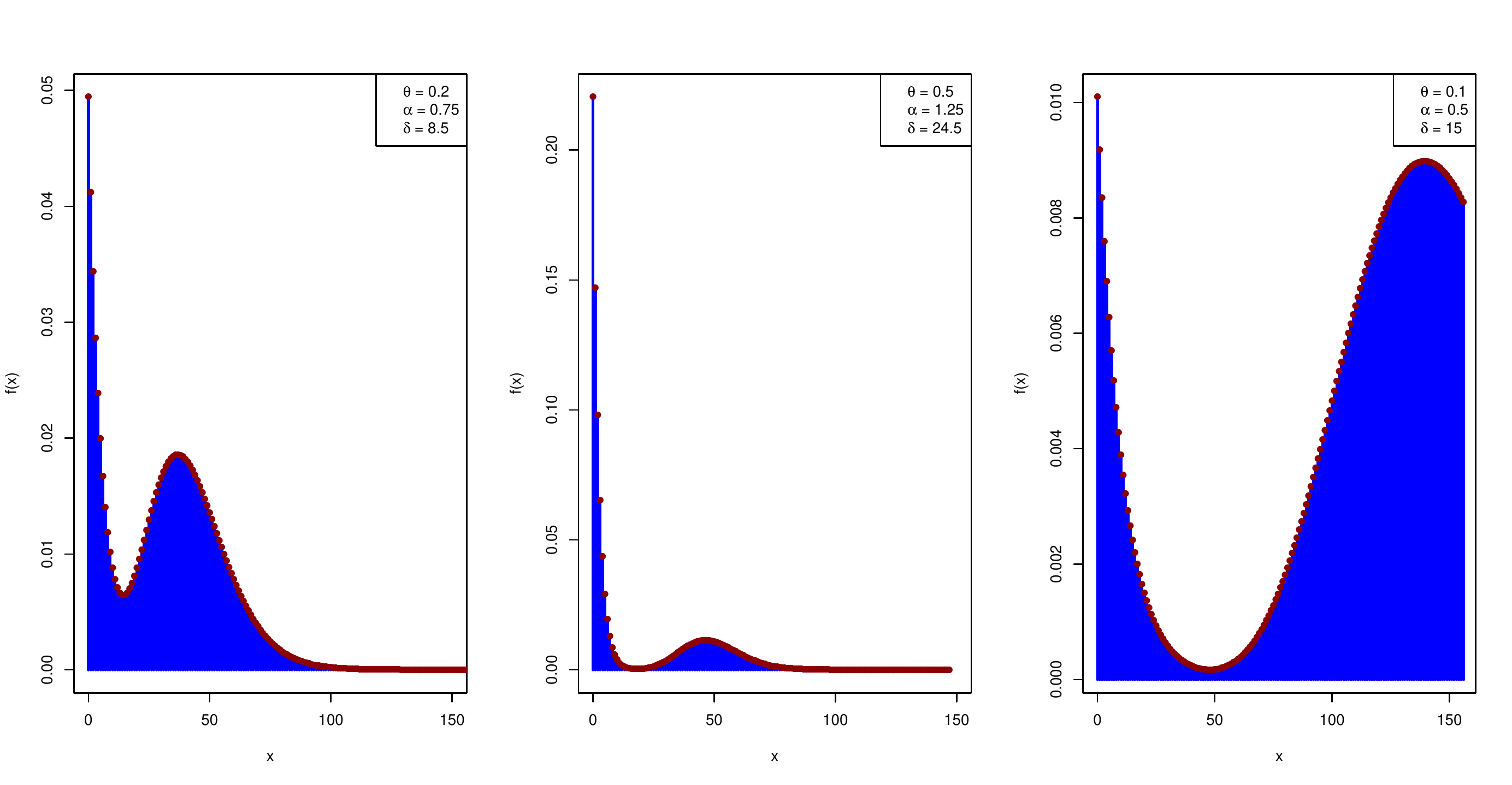}
   \caption{Some bimodal distributions of the PMQLD}
   \end{figure}
  \section{Statistical properties of PMQLD}
  In this section, we present some important statistical properties of the PMQLD such as the shape of the distribution,  conjugate prior for $\lambda$, moments and related measures, probability and moment generating functions, and quantile function. 
  \subsection{Shape of the distribution}
  From (6) we can easily derive $f(0)=\frac{\theta((1+\theta)^{\delta-1}\alpha^3+\theta^{\delta-1})}{(\alpha^3+1)(1+\theta)^\delta}$, and $\lim_{x \to\infty} f(x)=0.$\\\\
The recurrence relation for probabilities is given by
\begin{equation}
   \frac{f(x+1)}{f(x)}=\frac{A}{(x+1)(1+\theta)B};\hspace*{0.4cm} x=0,1,2,...,
\end{equation}
where $A=(1+\theta)^{\delta-1}\Gamma(\delta)\alpha^3\Gamma(x+2)+\theta^{\delta-1}(x+\delta)\Gamma(x+\delta)$, and\\\\
$B=(1+\theta)^{\delta-1}\Gamma(\delta)\alpha^3\Gamma(x+1)+\theta^{\delta-1}\Gamma(x+\delta)).$\\\\
The PMQLD$(\theta,\alpha,\delta)$ has a log-concave probability mass function when\begin{center}
$\Delta\eta(x)=\displaystyle\frac{f(x+1)}{f(x)}-\displaystyle\frac{f(x+2)}{f(x+1)}>0, \forall x$ (Gupta et al., 1997)
\end{center}\begin{center}
$\Rightarrow \displaystyle\frac{(x+2)A^2}{(x+1)\bigg((1+\theta)^{\delta-1}\Gamma(\delta)\alpha^3\Gamma(x+3)+\theta^{\delta-1}\Gamma(x+\delta+2)\bigg)B}>1$. \end{center} Under this condition, the distribution represents a unimodal distribution. Further, by using (8), it can be shown that
\begin{enumerate}[(i)]
\item For $(1+\theta)^{\delta-1}\alpha^3+\theta^{\delta-1}\delta<(1+\theta)((1+\theta)^{\delta-1}\alpha^3+\theta^{\delta-1}),$ equation (6) has unique mode at $X=0.$
\item Equation (6) has a unique mode at $X=x_{0},$ for \begin{center}
$\displaystyle\frac{(1+\theta)^{\delta-1}\Gamma(\delta)\alpha^3\Gamma(x_{0}+2)+\theta^{\delta-1}(x_{0}+\delta)\Gamma(x_{0}+\delta)}{(x_{0}+1)(1+\theta)\bigg((1+\theta)^{\delta-1}\Gamma(\delta)\alpha^3\Gamma(x_{0}+1)+\theta^{\delta-1}\Gamma(x_{0}+\delta)\bigg)}<1$, \end{center}and\begin{center}
$\displaystyle\frac{(1+\theta)^{\delta-1}\Gamma(\delta)\alpha^3\Gamma(x_{0})+\theta^{\delta-1}\Gamma(x_{0}+\delta-1)}{x_{0}(1+\theta)\bigg((1+\theta)^{\delta-1}\Gamma(\delta)\alpha^3\Gamma(x_{0}+1)+\theta^{\delta-1}\Gamma(x_{0}+\delta)\bigg)}<1.$\end{center}
\item Equation (6) has two modes at $X=x_{0}$ and $X=x_{0}+1$, for\begin{center} $(1+\theta)^{\delta-1}\Gamma(\delta)\alpha^3\Gamma(x_{0}+2)+\theta^{\delta-1}(x_{0}+\delta)\Gamma(x_{0}+\delta)$ \end{center}\begin{center}$=(x_{0}+1)(1+\theta)((1+\theta)^{\delta-1}\Gamma(\delta)\alpha^3\Gamma(x_{0}+1)+\theta^{\delta-1}\Gamma(x_{0}+\delta)).$\end{center}
\end{enumerate}
 The above facts are also shown in Figures 1-3 at different parameter settings.\\\\
Further, the PMQLD$(\theta,\alpha,\delta)$ has a log-convex probability mass function when $\Delta\eta(x)\leq0$:\begin{center}
$\Rightarrow \displaystyle\frac{(x+2)A^2}{(x+1)\bigg((1+\theta)^{\delta-1}\Gamma(\delta)\alpha^3\Gamma(x+3)+\theta^{\delta-1}\Gamma(x+\delta+2)\bigg)B}\leq1.$\end{center}
\subsection{Conjugate prior for $\lambda$}
In the PMQLD, the random variable $\Lambda$ of the Poisson parameter $\lambda$ is distributed as $\Lambda\sim $MQLD$(\theta,\alpha,\delta)$. Then, the posterior distribution $f(\lambda|x)$, given some datum or data $x$, is derived by the Bayes' theorem as
\begin{equation}
    f(\lambda|x)=\displaystyle\frac{f(x|\lambda)f(\lambda)}{\int_{0}^{\infty}f(x|\lambda)f(\lambda)d\lambda},
\end{equation}
where $f(x|\lambda)$ is the likelihood function. The following proposition provides the posterior distribution $f(\lambda|x)$.\\\\
\textbf{Proposition 2.} The posterior distribution $f(\lambda|x)$ is a two-component mixture of gamma $(\sum_{i=1}^{n}x_{i}+1,n+\theta)$ and gamma $(\sum_{i=1}^{n}x_{i}+\delta,n+\theta)$ with the mixing proportion\begin{center} $p=\displaystyle\frac{\alpha^3\Gamma(\delta)(n+\theta)^{\delta-1}\Gamma(\sum_{i=1}^{n}x_{i}+1)}{\alpha^3\Gamma(\delta)(n+\theta)^{\delta-1}\Gamma(\sum_{i=1}^{n}x_{i}+1)+\theta^{\delta-1}\Gamma(\sum_{i=1}^{n}x_{i}+\delta)}$, \end{center} which belongs to the same family of MQLD, i.e. $f(\lambda)$ is conjugate to $f(x|\lambda).$\\\\
\textbf{Proof:} see Appendix A
\subsection{Survival and hazard rate functions}
The survival/reliability function is associated with the probability of a system that will survive beyond a specified time. The survival function of the PMQLD is defined as
\begin{equation}
   S(x)=1-F(x) =1-\displaystyle\frac{\beta_{1}+\theta^\delta\Gamma(x+\delta+1)_{2}F_{1}(1,x+\delta+1;\delta+1;\frac{\theta}{1+\theta})}{\beta_{2}}
\end{equation}
where $\beta_{1}=\delta(1+\theta)^{\delta-1}\Gamma(\delta)\alpha^3\Gamma(x+1)((1+\theta)^{x+1}-1)$ and $\beta_{2}=(\alpha^3+1)\Gamma(\delta)x!\delta(1+\theta)^{x+\delta+1}.$
The hazard rate function (hrf) is the instantaneous failure rate. The hrf of the PMQLD is defined as
\begin{equation}
    h(x)=\lim_{\Delta x \to 0} \displaystyle\frac{P(x<X<x+\Delta x|X>x)}{\Delta x}=\displaystyle\frac{f(x)}{S(x)}=\frac{\delta\theta(1+\theta)\bigg((1+\theta)^{\delta-1}\Gamma(\delta)\alpha^3\Gamma(x+1)+\theta^{\delta-1}\Gamma(x+\delta)\bigg)}{\beta_{1}-(\beta_{2}+\theta^\delta\Gamma(x+\delta+1)_{2}F_{1}(1,x+\delta+1;\delta+1;\frac{\theta}{1+\theta}))}.
\end{equation}
Figure 4 provides an illustration of the possible shapes of the PMQLD's hazard rate function at different shape parameter values. According to these illustrations, it is clear that the proposed model has the capability to model the bathtub, monotonic increasing, and decreasing failure rate shapes.
\begin{figure}[h!]\label{fig:my_label}
        \includegraphics[width=17cm, height=10.5cm]{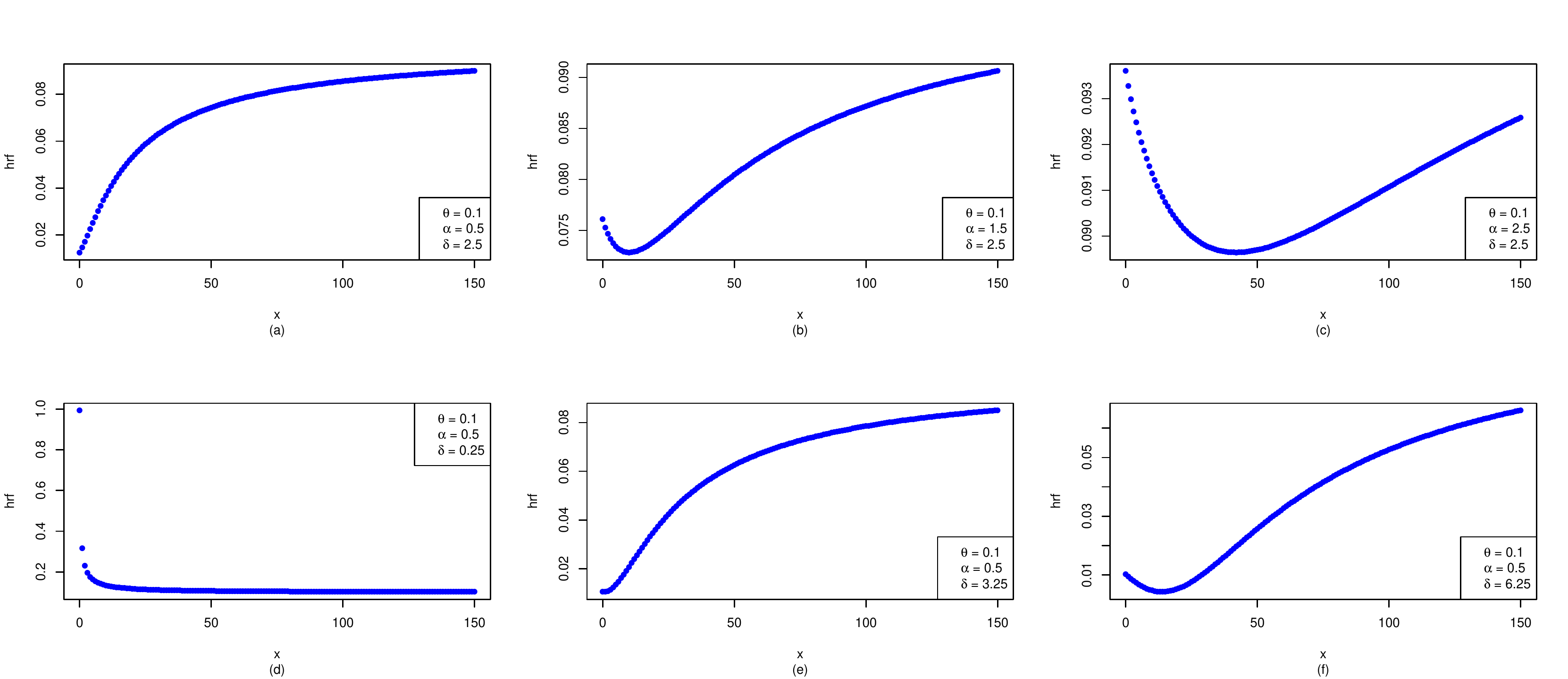}
   \caption{The hazard rate function at different values of $\alpha$, and $\delta$}
   \end{figure}
  \subsection{Moments and related measures}
  The central tendency, horizontal symmetry, tail heaviness, and dispersion are important characteristics of a distribution. These characteristics can be studied by using the moments. The following proposition provides the $r^{th}$ factorial moment of the PMQLD.\\\\
  \textbf{Proposition 3.} Let $X\sim$PMQLD$(\theta,\alpha,\delta)$, then the $r^{th}$ factorial moment of $X$ is given as
  \begin{equation}
      \mu_{(r)}^{'}=\displaystyle\frac{\Gamma(\delta)\Gamma(r+1)\alpha^3+\Gamma(\delta+r)}{(\alpha^3+1)\Gamma(\delta)\theta^r}.
  \end{equation}
  \textbf{Proof:}\begin{center}
  $\mu_{(r)}^{'}=E(\displaystyle\Pi_{i=0}^{r-1}(X-i))\hspace*{2cm}$\end{center}
  \begin{center}$=\displaystyle\sum_{x=0}^{\infty}x^{(r)}\displaystyle\int_{0}^{\infty} \displaystyle\frac{e^{-\lambda}\lambda^{x}}{x!}\displaystyle\frac{\theta e^{-\theta \lambda}}{(\alpha^3+1)\Gamma(\delta)}\bigg(\Gamma(\delta)\alpha^3 + (\theta \lambda)^{\delta-1}\bigg)d\lambda$\end{center}
  \begin{center}$=\displaystyle\int_{0}^{\infty}\lambda^{r} \displaystyle\frac{\theta e^{-\theta \lambda}}{(\alpha^3+1)\Gamma(\delta)}\bigg(\Gamma(\delta)\alpha^3 + (\theta \lambda)^{\delta-1}\bigg)d\lambda$\end{center}
  \begin{center}$=\displaystyle\frac{\theta}{(\alpha^3+1)\Gamma(\delta)}\bigg(\displaystyle\frac{\Gamma(\delta)\alpha^3\Gamma(r+1)}{\theta^{r+1}}+\displaystyle\frac{\Gamma(\delta+r)}{\theta^{r+1}}\bigg)$\end{center}
   \begin{center}$=\displaystyle\frac{\Gamma(\delta)\Gamma(r+1)\alpha^3+\Gamma(\delta+r)}{(\alpha^3+1)\Gamma(\delta)\theta^r}$.\end{center}
   Then, the first four raw moments of $X$ can be derived by the following relationship\begin{center}
   $\mu_{r}^{'}=E(x^r)=\displaystyle\sum_{i=0}^{r}S(r,i)\mu_{(i)}^{'}\hspace*{0.15cm};\hspace*{0.15cm} r=1,2,...,$\end{center}
   where $S(r,i)$ is the Stirling numbers of the second kind, and it is defined as\begin{center}
   $S(r,i)=\displaystyle\frac{1}{i!}\displaystyle\sum_{j=0}^{i}(-1)^{i-j}{i\choose j}j^r\hspace*{0.15cm} ,\hspace*{0.15cm} 0<i<r$.\end{center}
   Let $\kappa_{1}=\alpha^3+\delta, \hspace*{0.1cm}\kappa_{2}=2\alpha^3+\delta(\delta+1),\hspace*{0.1cm}\kappa_{3}=6\alpha^3+\delta(\delta+1)(\delta+2),$ and $\kappa_{4}=24\alpha^3+\delta(\delta+1)(\delta+2)(\delta+3).$ Then,\\\\ $\mu_{1}^{'}=\displaystyle\frac{\kappa_{1}}{(\alpha^3+1)\theta}=\mu\hspace*{0.05cm}, \hspace*{0.5cm}\mu_{2}^{'}=\displaystyle\frac{\theta\kappa_{1}+\kappa_{2}}{(\alpha^3+1)\theta^2},$
   $\hspace*{0.5cm}\mu_{3}^{'}=\displaystyle\frac{\theta^2\kappa_{1}+3\theta\kappa_{2}+\kappa_{3}}{(\alpha^3+1)\theta^3},$ and
   $\mu_{4}^{'}=\displaystyle\frac{\theta^3\kappa_{1}+7\theta^2\kappa_{2}+6\theta\kappa_{3}+\kappa_{4}}{(\alpha^3+1)\theta^4}$.\\\\
   Further, the $r^{th}$-order moments about the mean can be obtained by using the relationship between moments about the mean and moments about the origin, i.e. \begin{center}$\mu_{r}=E\Bigg[(Y-\mu)^r\Bigg]=\displaystyle\sum_{i=0}^{r}{r\choose i}(-1)^{r-i}\mu_{i}^{'}\mu^{r-i};\hspace*{0.2cm} r=1,2,...$.\end{center}Therefore, the variance of $X, \sigma^2$ and index of dispersion, $\gamma_{1}$ are derived as\begin{center}
   $\sigma^2=\mu_{2}=\displaystyle\frac{\alpha^6(1+\theta)+\alpha^3(2+\theta+\delta(\theta+\delta-1))+\delta(\theta+1)}{(\alpha^3+1)^2\theta^2}=\mu+\mu^2\bigg(\frac{\alpha^3(\alpha^3+2+\delta(\delta-1))+\delta}{(\alpha^3+\delta)^2}\bigg)$, \end{center}and\begin{center}
   $\gamma_{1}=\displaystyle\frac{\mu_{2}}{\mu_{1}^{'}}=1+\displaystyle\frac{\alpha^3(\alpha^3+2+\delta(\delta-1))+\delta}{(\alpha^3+1)(\alpha^3+\delta)\theta}$,\end{center} respectively. It is clear that the $\gamma_{1}>1$. Then, the PMQLD is an over-dispersed distribution.
   Since the mathematical expressions of $\mu_{3}$ and $\mu_{4}$ are very long, we present the graphical presentations of the skewness  $(\gamma_{2})=\displaystyle\frac{\mu_{3}}{(\mu_{2})^{3/2}}$ and kurtosis $(\gamma_{3})=\displaystyle\frac{\mu_{4}}{\mu_{2}^{2}}$ of the PMQLD in Figure 6.
   The surface plots in Figures 5, and 6 show some possible values of the index of dispersion, skewness, and kurtosis that can be accommodated by the PMQLD at different settings of the parameters. Hence, these plots indicate that the PMQLD $(\theta,\alpha,\delta)$ has the capability to accommodate various ranges of the index of dispersion, skewness, and kurtosis at different sets of parameters for over-dispersed count data. 
   \begin{figure}[h!]\label{fig:my_label}
        \includegraphics[width=17cm, height=10.0cm]{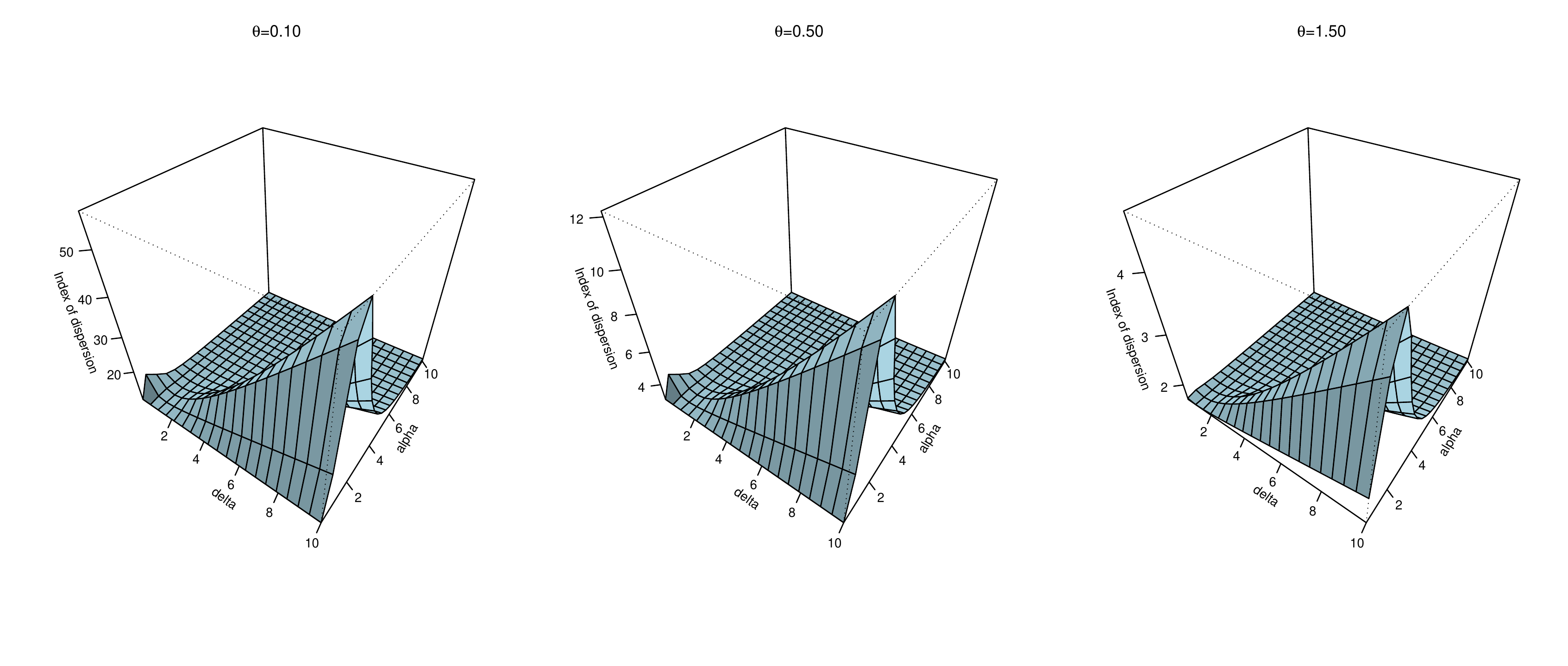}
   \caption{Surface plots for the index of dispersion at different values of $\theta$, $\alpha$, and $\delta$}
   \end{figure}
   \begin{figure}[h!]\label{fig:my_label}
        \includegraphics[width=14cm, height=8.59cm]{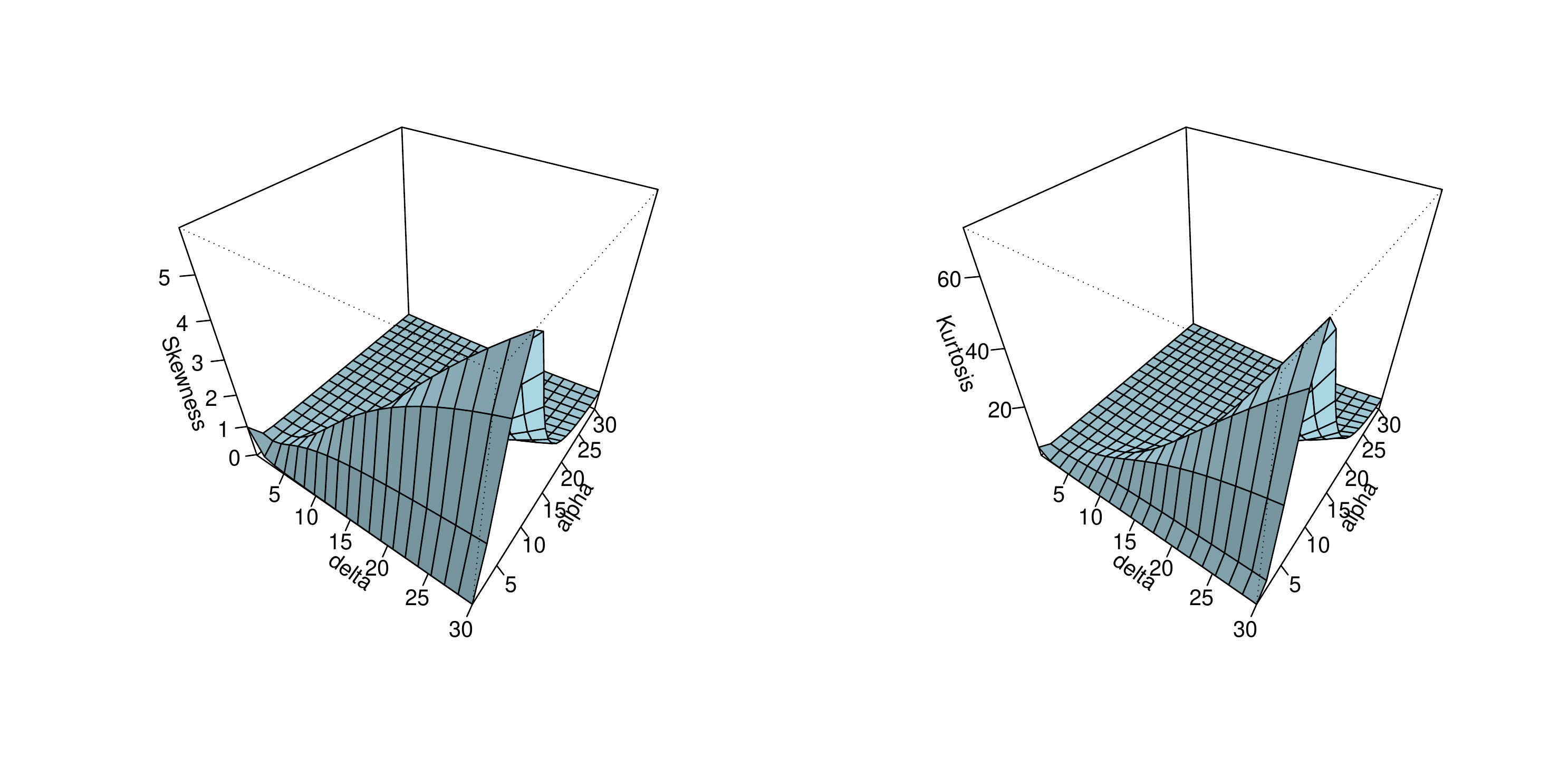}
   \caption{Surface plots for the skewness and kurtosis functions at different values of $\theta$, and $\alpha$}
   \end{figure}
   \subsection{Probability and moment generating functions}
   The characteristics of a probability distribution are directly associated with its probability generating function (pgf) and the moment generating function (mgf). The following proposition provides the pgf of the PMQLD.\\\\
   \textbf{Proposition 4.} The pgf, $G(t)=E(t^X),$ $X\sim$PMQLD$(\theta,\alpha,\delta)$ is given as
   \begin{equation}
     G(t)=\displaystyle\frac{\theta(\alpha^3(1-t+\theta)^{\delta-1}+\theta^{\delta-1})}{(\alpha^3+1)(1-t+\theta)^\delta} , t\in\mathbb{R}.  
   \end{equation}
   \textbf{Proof:} \begin{center}$G(t)=E(t^{X})$\end{center}
   \begin{center}$=\displaystyle\sum_{x=0}^{\infty}t^{X}\displaystyle\int_{0}^{\infty} \displaystyle\frac{e^{-\lambda}\lambda^{x}}{x!}\displaystyle\frac{\theta e^{-\theta \lambda}}{(\alpha^3+1)\Gamma(\delta)}\bigg(\Gamma(\delta)\alpha^3 + (\theta \lambda)^{\delta-1}\bigg)d\lambda$\end{center}\begin{center}$=\displaystyle\int_{0}^{\infty}e^{-\lambda(1-t)}\displaystyle\frac{\theta e^{-\theta \lambda}}{(\alpha^3+1)\Gamma(\delta)}\bigg(\Gamma(\delta)\alpha^3 + (\theta \lambda)^{\delta-1}\bigg)d\lambda$\end{center}
   \begin{center}$=\displaystyle\frac{\theta}{(\alpha^3+1)\Gamma(\delta)}\bigg(\Gamma(\delta)\alpha^3\displaystyle\int_{0}^{\infty}e^{-\lambda(1-t+\theta)}d\lambda+\theta^{\delta-1}\displaystyle\int_{0}^{\infty}e^{-\lambda(1-t+\theta)}\lambda^{\delta-1}d\lambda\bigg)$\end{center}
   \begin{center}$=\displaystyle\frac{\theta(\alpha^3(1-t+\theta)^{\delta-1}+\theta^{\delta-1})}{(\alpha^3+1)(1-t+\theta)^\delta}$.\end{center}
   The mgf can be obtained effortlessly from pgf by using the relationship $G(e^t)=E(e^{tX})=M_{X}(t)$, and given as
   \begin{equation}
      M_{X}(t)=\displaystyle\frac{\theta(\alpha^3(1-e^t+\theta)^{\delta-1}+\theta^{\delta-1})}{(\alpha^3+1)(1-e^t+\theta)^\delta}, t\in\mathbb{R}.  
   \end{equation}
   \subsection{Quantile function}
 The quantile function is a useful function to estimate the quantiles. Let us define the quantiles for random variable $X\sim$PMQLD$(\theta,\alpha,\delta)$. The $u^{th}$ quantile can be derived by solving $F(x_{u})=u$ for $x_{u}, 0<u<1.$ Then, the $u^{th}$ quantile function of the PMQLD is given as
 \begin{equation}
   \beta_{1}(x_{u})+\theta^\delta\Gamma(x_{u}+\delta+1)_{2}F_{1}(1,x_{u}+\delta+1;\delta+1;\frac{\theta}{1+\theta})-u\beta_{2}(x_{u})=0, \hspace*{0.2cm}0<u<1, 
 \end{equation}
 where $\beta_{1}(x_{u})=\delta(1+\theta)^{\delta-1}\Gamma(\delta)\alpha^3\Gamma(x_{u}+1)((1+\theta)^{x_{u}+1}-1)$ and $\beta_{2}(x_{u})=(\alpha^3+1)\Gamma(\delta)x_{u}!\delta(1+\theta)^{x_{u}+\delta+1}.$\\\\
 Since equation (15) is not a closed-form in $x_{u}$, the estimates of the quantiles can be evaluated by using any numerical method. Further, the first three quartiles can be calculated by substituting $u=0.25, 0.50,$ and $0.75$ in equation (15) and solving the respective equations.
  \section{Zero-modified PMQLD (ZMPMQLD)}
  The actual count data distribution is modified at zero probability to adopt the situation with an excessive number of zeros or a smaller number of zeros than those from the actual count data distribution. \\\\The random variable $X$ is said to have the Zero-modified PMQLD (ZMPMQLD) $(\phi,\theta,\alpha,\delta)$ if its pmf is given as
  \[f(x)=
\begin{cases}
\phi+(1-\phi)\displaystyle\frac{\theta((1+\theta)^{\delta-1}\alpha^3+\theta^{\delta-1})}{(\alpha^3+1)(1+\theta)^\delta}&\text{for $x=0$}\\
(1-\phi)\displaystyle\frac{\theta\bigg(\Gamma(\delta)\Gamma(x+1)\alpha^3(1+\theta)^{\delta-1}+\theta^{\delta-1}\Gamma(x+\delta)\bigg)}{x !(\alpha^3+1)  (1+\theta)^{x+\delta}\Gamma(\delta)}&\text{for $x=1,2,...$}\\
\end{cases}
\]
  \begin{equation}
  \end{equation}
where $\delta>0, \theta>0, \alpha^3>-1$, and $\frac{\theta((1+\theta)^{\delta-1}\alpha^3+\theta^{\delta-1})}{\theta^\delta-(1+\theta)^{\delta-1}(1+\theta+\alpha^3)}\leq\phi\leq1$. The parameter $\phi$ is the zero-modified parameter. Note that equation (16) is not a finite mixture model since $\phi$ can take negative values. Further, various $\phi$ values adopt various zero-modifications of the PMQLD.
\begin{enumerate}[(i)]
  \item When $\phi=\frac{\theta((1+\theta)^{\delta-1}\alpha^3+\theta^{\delta-1})}{\theta^\delta-(1+\theta)^{\delta-1}(1+\theta+\alpha^3)}$, the ZMPMQLD reduces to the zero-truncated PMQLD. Here, $\phi$ is no longer appears. The zero-truncated models are commonly used to study the length of hospital stay.
  \item For $\frac{\theta((1+\theta)^{\delta-1}\alpha^3+\theta^{\delta-1})}{\theta^\delta-(1+\theta)^{\delta-1}(1+\theta+\alpha^3)}<\phi<0$, the ZMPMQLD reduces to the zero-deflated PMQLD, and the zero-deflated models are very rare in practice.
  \item When $\phi=0$, the ZMPMQLD is the PMQLD.
  \item For $0<\phi<1$, the ZMPMQLD reduces to the zero-inflated PMQLD. This can accommodate more zeros than the actual PMQLD.
  \item When $\phi=1$, the ZMPMQLD is degenerated at zero, i.e. all probabilities of the distribution are concentrated at zero.
\end{enumerate}
\section{Simulation and parameter estimation}
\subsection{Simulation of the random variables}
Here, we provide two different algorithms to simulate the random variables $x_{1}, x_{2}, . . . ,x_{n}$ from the PMQLD$(\theta,\alpha,\delta)$ with size $n$ based on the inverse transform method. \\\\The first algorithm is obtained by considering the mixing of the PMQLD. Since $X|\Lambda\sim$Poisson $(\lambda)$ and $\Lambda\sim$MQLD$(\theta,\alpha,\delta)$, the first algorithm is obtained as follows\\\\
\textbf{Algorithm I:}
\begin{enumerate}[i]
\item Simulate the random variables, $u_{i}\sim$uniform$(0,1);\hspace*{0.2cm}i=1,2,...,n$.
\item Solve the non-linear equation for $\lambda_{i}$:
$\Gamma(\delta)(1+\alpha^3(1-e^{-\theta\lambda_{i}}))-\Gamma(\delta,\theta \lambda_{i})-u_{i}(\alpha^3+1)\Gamma(\delta)=0$ to simulate the random variables, $\lambda_{i}\sim$MQLD$(\theta,\alpha,\delta);\hspace*{0.2cm}i=1,2,...,n$.
\item Simulate $x_{i}$ from Poisson $(\lambda_{i});\hspace*{0.2cm}i=1,2,...,n$.
\end{enumerate}
The second algorithm is obtained from the quantile function of PMQLD discussed in subsection 3.6, and the steps are as follows\\\\
\textbf{Algorithm II:}
\begin{enumerate}[i]
\item Simulate the random variables, $u_{i}\sim$uniform$(0,1);\hspace*{0.1cm}i=1,2,...,n$.
\item Solve the non-linear equation for $[x_{u_{i}}];\\\hspace*{0.1cm}  \beta_{1}(x_{u_{i}})+\theta^\delta\Gamma(x_{u_{i}}+\delta+1)_{2}F_{1}(1,x_{u_{i}}+\delta+1;\delta+1;\frac{\theta}{1+\theta})-u_{i}\beta_{2}(x_{u_{i}})=0,$
where $\beta_{1}(x_{u_{i}})$ and $\beta_{2}(x_{u_{i}})$ are defined as in section 3.6. $[.]$ denotes the integer part.
\end{enumerate}
\subsection{Parameter estimation of PMQLD}
In this subsection, we discuss the parameter estimation of the PMQLD by using the method of moment estimation and the maximum likelihood estimation method.
\subsubsection{Method of moment estimation (MME)}
Given a random samples $x_{1},x_{2}...x_{n}$ with size $n$ from the PMQLD$(\theta,\alpha,\delta)$, the method of moment estimators of $\theta,\alpha,$ and $\delta$, abbreviated as $\hat{\theta}_{MME}, \hat{\alpha}_{MME},$ and $\hat{\delta}_{MME},$ can be evaluated by equating the raw-moments, say $\mu_{r}^{'}$, to the sample moments, say $\displaystyle\frac{\displaystyle\sum_{i=1}^{n} x_{i}^{r}}{n}, r=1, 2, 3$, i.e. we need to find the solutions of the following system of non-linear equations:\\\\
$n\kappa_{1}-(\alpha^3+1)\theta\sum_{i=1}^{n}x_{i}=0;\hspace*{0.1cm} n(\theta\kappa_{1}+\kappa_{2})-(\alpha^3+1)\theta^2\sum_{i=1}^{n}x_{i}^{2}=0;\hspace*{0.1cm} n(\theta^2\kappa_{1}+3\kappa_{2}+\kappa_{3})-(\alpha^3+1)\theta^3\sum_{i=1}^{n}x_{i}^{3}=0,$\\\\
where $\kappa_{1},\kappa_{2},$ and $\kappa_{3}$ are defined in subsection 3.4. It is clear that these equations are not a closed form. However, the solutions can be derived by using a numerical method.
\subsubsection{Maximum likelihood estimation (MLE)}
Given a random samples $x_{1},x_{2}...x_{n}$ with size $n$ from the PMQLD$(\theta,\alpha,\delta)$, the likelihood function of the $i^{th}$ sample value $x_{i}$ is given as\begin{center}$
L(\theta,\alpha,\delta|x)=
\displaystyle\frac{\theta}{x_{i} !(\alpha^3+1)  (1+\theta)^{x_{i}+\delta}\Gamma(\delta)}\bigg(\Gamma(\delta)\Gamma(x_{i}+1)\alpha^3(1+\theta)^{\delta-1}+\theta^{\delta-1}\Gamma(x_{i}+\delta)\bigg) $.\end{center}
Then, the log-likelihood function is given as\begin{center}$
log(L(\theta,\alpha,\delta|x_{i}))=l(\theta,\alpha,\delta|x)$\end{center}\begin{center}$=n\bigg(log(\theta)-log(\alpha^3+1)-log(\Gamma(\delta))\bigg)+\sum_{i=1}^{n}log\bigg(\Gamma(\delta)\Gamma(x_{i}+1)\alpha^3(1+\theta)^{\delta-1}+\theta^{\delta-1}\Gamma(x_{i}+\delta)\bigg)$\end{center}\begin{equation}-log(x_{i}!)-(x_{i}+\delta)log(1+\theta).
\end{equation}
The score functions are\begin{center}
$\displaystyle\frac{\partial l(\theta,\alpha,\delta|x)}{\partial\theta}=\frac{n}{\theta}+\sum_{i=1}^{n}\frac{\Gamma(\delta)\Gamma(x_{i}+1)\alpha^3(\delta-1)(1+\theta)^{\delta-2}+\Gamma(x_{i}+\delta)(\delta-1)\theta^{\delta-2}}{\Gamma(\delta)\Gamma(x_{i}+1)\alpha^3(1+\theta)^{\delta-1}+\theta^{\delta-1}\Gamma(x_{i}+\delta)}-\sum_{i=1}^{n}\frac{x_{i}+\delta}{1+\theta},$\end{center}
\begin{center}$\displaystyle\frac{\partial l(\theta,\alpha,\delta|x)}{\partial\alpha}=\sum_{i=1}^{n}\frac{3\alpha^2\Gamma(\delta)\Gamma(x_{i}+1)(1+\theta)^{\delta-1}}{\Gamma(\delta)\Gamma(x_{i}+1)\alpha^3(1+\theta)^{\delta-1}+\theta^{\delta-1}\Gamma(x_{i}+\delta)}-\frac{3n\alpha^2}{\alpha^3+1},$ \end{center} and \begin{center}
$\displaystyle\frac{\partial l(\theta,\alpha,\delta|x)}{\partial\delta}=\sum_{i=1}^{n}\frac{\Gamma(x_{i}+1)\alpha^3(\Gamma(\delta)(1+\theta)^{\delta-1}log(1+\theta)+(1+\theta)^{\delta-1}\Gamma(\delta)\psi(\delta))+\Gamma(x_{i}+\delta)\theta^{\delta-1}(\log(\theta)+\psi(x_{i}+\delta))}{\Gamma(\delta)\Gamma(x_{i}+1)\alpha^3(1+\theta)^{\delta-1}+\theta^{\delta-1}\Gamma(x_{i}+\delta)}$\end{center}\begin{center}$-n(log(1+\theta)+\psi(\delta)),$\end{center}
where $\psi(a)=\displaystyle\frac{\partial}{\partial a}log \Gamma(a)=\frac{\Gamma^{'}(a)}{\Gamma(a)}$. By setting the score functions equal to zero, the maximum likelihood estimators of $\theta,\alpha,$ and $\delta$, abbreviated as $\hat{\theta}_{MLE}, \hat{\alpha}_{MLE},$ and $\hat{\delta}_{MLE}$ can be derived. These systems of non-linear equations can be solved by a numerical method. Here,  the solutions of the parameter estimates will be obtained by using the $optim$ function in the R package $stats$. \\\\The asymptotic confidence intervals for the parameters $\theta, \alpha,$ and $\delta$ are derived by the asymptotic theory. The estimators are asymptotic three-variate normal with mean $(\theta,\alpha, \delta)$ and the observed information matrix
 \[ I(\theta, \alpha, \delta)= \left( \begin{array}{ccc}
-\displaystyle\frac{\partial^{2} l(\theta, \alpha, \delta|x)}{\partial \theta^{2}}&-\displaystyle\frac{\partial^{2} l(\theta, \alpha, \delta|x)}{\partial \theta \partial \alpha}&-\displaystyle\frac{\partial^{2} l(\theta, \alpha, \delta|x)}{\partial \theta \partial \delta}\\
-\displaystyle\frac{\partial^{2} l(\theta, \alpha, \delta|x)}{\partial \alpha \partial \theta}&-\displaystyle\frac{\partial^{2} l(\theta, \alpha, \delta|x)}{ \partial \alpha^2}&-\displaystyle\frac{\partial^{2} l(\theta, \alpha, \delta|x)}{\partial \alpha \partial \delta}\\
-\displaystyle\frac{\partial^{2} l(\theta, \alpha, \delta|x)}{\partial \delta \partial \theta}&-\displaystyle\frac{\partial^{2} l(\theta, \alpha, \delta|x)}{\partial \delta \partial \alpha}&-\displaystyle\frac{\partial^{2} l(\theta, \alpha, \delta|x)}{\partial \delta^2}\\
\end{array} \right)\] 
at $\theta=\hat{\theta}_{MLE}, \alpha=\hat{\alpha}_{MLE},$ and $\delta=\hat{\delta}_{MLE} $,  i.e. $(\hat{\theta}_{MLE}, \hat{\alpha}_{MLE}, \hat{\delta}_{MLE}) \sim N_{3}((\theta,\alpha,\delta), I^{-1}(\theta,\alpha,\delta)).$ The second order partial derivatives of the log-likelihood function are given in Appendix B.\\\\
Therefore, a $(1-a)100\%$ confidence interval for the parameters $\theta, \alpha,$ and $\delta$ are given by\begin{center}$\hat{\theta_{MLE}}\pm z_{a/2}\sqrt{Var(\hat{\theta_{MLE}})}    ,\hspace*{0.5cm}
\hat{\alpha}_{MLE}\pm z_{a/2}\sqrt{Var(\hat{\alpha}_{MLE})}  ,\hspace*{0.5cm}$
  $\hat{\delta}_{MLE}\pm z_{a/2}\sqrt{Var(\hat{\delta}_{MLE})},\hspace*{0.5cm}$
  \end{center} wherein, the $Var(\hat{\theta}_{MLE}), Var(\hat{\alpha}_{MLE}),$ and $Var(\hat{\delta}_{MLE})$ are the variance of $\hat{\theta}_{MLE}, \hat{\alpha}_{MLE},$ and $\hat{\delta}_{MLE}$, respectively, and can be derived by diagonal elements of $I^{-1}(\theta, \alpha, \delta)$ and $z_{a/2}$ is the critical value at $a$ level of significance.
\subsection{Parameter estimation of ZMPMQLD}
Here, we discuss the parameter estimation of ZMPMQLD by using the maximum likelihood estimation method.\\\\
Given a random sample $x_{1},x_{2}...x_{n}$ with size $n$ from the ZMPMQLD$(\phi,\theta,\alpha,\delta)$, the likelihood function of the $i^{th}$ sample value $x_{i}$ is given as\begin{center}$L(\phi,\theta,\alpha,\delta|x_{i})=\bigg(\phi+(1-\phi)\displaystyle\frac{\theta((1+\theta)^{\delta-1}\alpha^3+\theta^{\delta-1})}{(\alpha^3+1)(1+\theta)^\delta}\bigg)^{I_{(X=0)}(x_{i})}\times$\end{center}\begin{center}$\bigg((1-\phi)\displaystyle\frac{\theta(\Gamma(\delta)\Gamma(x_{i}+1)\alpha^3(1+\theta)^{\delta-1}+\theta^{\delta-1}\Gamma(x_{i}+\delta)}{x_{i}!(\alpha^3+1)(1+\theta)^{x_{i}+\delta}\Gamma(\delta)}\bigg)^{(1-I_{(X=0)}(x_{i}))},$\end{center}
where $I_{S}(.)$ is the indicator function of subset $S$. Then, the log-likelihood function is given as\begin{center}
$l(\phi,\theta,\alpha,\delta|x)=\displaystyle n_{0}log\bigg(\phi+(1-\phi)\frac{\theta((1+\theta)^{\delta-1}\alpha^3+\theta^{\delta-1})}{(\alpha^3+1)(1+\theta)^\delta}\bigg)+(n-n_{0})log\bigg(\frac{(1-\phi)\theta}{(\alpha^3+1)\Gamma(\delta)}\bigg)+$\end{center}\begin{center}$\displaystyle\sum_{i=1}^{n}(1-I_{(X=0)}(x_{i}))\Bigg(log\bigg(\Gamma(\delta)\Gamma(x_{i}+1)\alpha^3(1+\theta)^{\delta-1}+\theta^{\delta-1}\Gamma(x_{i}+\delta)\bigg)-log\bigg(x_{i}!(1+\theta)^{x_{i}+\delta}\bigg)\Bigg).$\end{center}
The score functions of the above log-likelihood function are given in Appendix C. The maximum likelihood estimators of $\phi, \theta,\alpha,$ and $\delta$, and inferences can be derived in the similar way that is mentioned in subsection 5.2.2.
\section{Monte Carlo simulation study and real-world application}
This section is devoted to discuss the simulation study and the applicability of PMQLD.
\subsection{Monte Carlo simulation study}
Here, we examine the accuracy of the MLE method in the unknown parameter estimation of the PMQLD with respect to sample size $n$. The second algorithm given in subsection 5.1 is used to simulate the random variables from the PMQLD. The sample sizes are taken as $60,100,200,$ and $300$, and the simulation study is repeated 1000 times. The study is designed as follows
\begin{enumerate}[(i)]
\item Simulate 1000 samples of size $n$. 
\item Compute the maximum likelihood estimates for the 1000 samples, say $(\hat{\theta}_{i}, \hat{\alpha}_{i}, \hat{\delta}_{i}), i=1,2,...1000.$
\item Compute the average MLEs, biases, and mean square errors (MSEs) by using the following equations\begin{center}
$\hat{s}(n)=\frac{1}{1000}\sum_{i=1}^{1000}\hat{s}_{i}, \hspace*{0.3cm}bias_{s}(n)=\frac{1}{1000}\sum_{i=1}^{1000}(\hat{s}_{i}-s),$ and $MSE_{s}(n)=\frac{1}{1000}\sum_{i=1}^{1000}(\hat{s}_{i}-s)^2,$\end{center}
for $s=\theta, \alpha, \delta,$ and $n=60,100,200,300$.
\end{enumerate}
Tables 2 to 5 represent the average MLEs, biases, and MSEs (in parentheses) of $\theta,\alpha,$ and $\delta$ for different values of $\theta,\alpha,$ and $\delta$ which are $\theta=0.1$, $0.3$; $\alpha=0.25,$ $0.50$, $0.75$; and $\delta=2.50,$ $3.50$, $4.50$. Note that the biases and MSEs decrease as $n$ increases for all parameters. Then, MLE method verifies the asymptotic property for all parameter estimates, and the parameters $\theta, \alpha,$ and $\delta$ are over estimated. Further, while the estimation of $\theta$ is good for small value of $\theta,$ the estimation of $\alpha$ doses not show a good estimation for small value of $\alpha$ based on average biases and MSEs. However, there is no particular pattern for estimation of $\delta.$\newpage
	Table 2. Performance of MLE method for the PMQLD$(\theta=0.10,\alpha=0.50,\delta)$
\begin{table}[H]
\centering
\begin{small}
  \begin{tabular}{ c|cccccccc} 
  \hline
  &\multicolumn{2}{c}{$n=60$}&\multicolumn{2}{c}{$n=100$}&\multicolumn{2}{c}{$n=200$}&\multicolumn{2}{c}{$n=300$}\\
  \hline
  &MLE&Bias(MSE)&MLE&Bias(MSE)&MLE&Bias(MSE)&MLE&Bias(MSE)\\
  \hline
  $\delta=2.50$&&&&&&&&\\
  $\theta$&0.1151&0.0152(0.0014)&0.1086&0.0086(0.0007)&0.1067&	0.0067(0.0002)&	0.1009&	0.0009(0.0001)\\
  $\alpha$&0.7519&	0.2519(0.0691)&	0.7482&	0.2482(0.0648)&	0.7290&	0.2290(0.0557)&	0.7195&	0.2195(0.0494)\\
  $\delta$&3.0685&	0.5685(1.0107)&3.0545&	0.5545 (0.9155)&	2.8466&	0.3466(0.3078)&	2.8358&	0.3358(0.2186)\\
\hline
$\delta=3.50$&&&&&&&&\\
$\theta$&0.1101&	0.0101(0.0010)&	0.1051&	0.0051(0.0006)&	0.1044&	0.0044(0.0002)&	0.1029&	0.0029(0.0001)\\
$\alpha$&0.6733&	0.1733(0.0348)&	0.6662&	0.1662(0.0317)&	0.6684	&0.1684(0.0300)&	0.6509&	0.1509(0.0259)\\
$\delta$&3.8991	&0.3991(0.9110)	&3.8494&	0.3494(0.8928)&	3.6791&	0.1791(0.2898)&	3.6010&	0.1010(0.0887)\\
\hline
$\delta=4.50$&&&&&&&&\\
$\theta$&0.1087&	0.0087(0.0009)&	0.1054&	0.0054(0.0006)&	0.1037&	0.0037(0.0002)&	0.1029&	0.0021(0.0001)\\
$\alpha$&0.6299	&0.1299(0.0213)&	0.6268&	0.1268(0.0191)&	0.6217&	0.1217(0.0161)&	0.6192&	0.1192(0.0156)\\
$\delta$&4.8664	&0.3664(1.3188)&	4.7506&	0.2506(1.1957)&	4.6354&	0.1354(0.3871)&	4.5805&	0.0805(0.0122)\\
\hline
\end{tabular}
\end{small}
\end{table}
Table 3. Performance of MLE method for PMQLD$(\theta=0.30,\alpha=0.50,\delta)$
\begin{table}[H]
\centering
\begin{small}
  \begin{tabular}{ c|cccccccc} 
  \hline
  &\multicolumn{2}{c}{$n=60$}&\multicolumn{2}{c}{$n=100$}&\multicolumn{2}{c}{$n=200$}&\multicolumn{2}{c}{$n=300$}\\
  \hline
  &MLE&Bias(MSE)&MLE&Bias(MSE)&MLE&Bias(MSE)&MLE&Bias(MSE)\\
  \hline
  $\delta=2.50$&&&&&&&&\\
  $\theta$&0.3791&	0.0790 (0.0242)&	0.3635&	0.0635(0.0215)	&0.3378	&0.0378 (0.0075)&	0.3367&	0.0367(0.0039)\\
  $\alpha$&0.8392&	0.3392 (0.1199)&	0.8243&	0.3243(0.1109)	&0.7956&	0.2956 (0.0933)&	0.7503&	0.2503(0.0853)\\
  $\delta$&3.6229&	1.1229(3.5560)&	3.4672&	0.9672(2.9115)	&3.1876&	0.6876(1.2228)&	3.1509&	0.6509 (0.7488)\\
\hline
$\delta=3.50$&&&&&&&&\\
$\theta$&0.3839	&0.0839 (0.0245)&	0.3710&	0.0710(0.0216)&	0.3275&	0.0275(0.0042)&	0.3207&	0.0207(0.0018)\\
$\alpha$&0.7493	&0.2493 (0.0698)&	0.7215&	0.2215(0.0543)&	0.7170&	0.2170(0.0487)&	0.6983&	0.1983(0.0457)\\
$\delta$&4.3284	&0.8284(2.5256)&	4.1190&	0.6190(1.9316)&	3.7680&	0.2680(0.4910)&	3.7393&	0.2393(0.2547)\\
\hline
$\delta=4.50$&&&&&&&&\\
$\theta$&0.3638&0.0638 (0.0210)	&0.3484&	0.0484(0.0118)&	0.3143&	0.0143(0.0033)&	0.3117&	0.0117(0.0016)\\
$\alpha$&0.6784	&0.1784 (0.0373)&	0.6703&	0.1703(0.0330)&	0.6658&	0.1658(0.0299)&	0.6578&	0.1578(0.0277)\\
$\delta$&5.1958&	0.6958(3.0589)&	4.9240&	0.4240(2.0855)&	4.6154&	0.1154(0.6430)&	4.6713&	0.1713(0.2667)\\
\hline
\end{tabular}
\end{small}
\end{table}
Table 4. Performance of MLE method for PMQLD$(\theta=0.10,\alpha,\delta=2.50)$
\begin{table}[H]
\centering
\begin{small}
  \begin{tabular}{ c|cccccccc} 
  \hline
  &\multicolumn{2}{c}{$n=60$}&\multicolumn{2}{c}{$n=100$}&\multicolumn{2}{c}{$n=200$}&\multicolumn{2}{c}{$n=300$}\\
  \hline
  &MLE&Bias(MSE)&MLE&Bias(MSE)&MLE&Bias(MSE)&MLE&Bias(MSE)\\
  \hline
  $\alpha=0.25$&&&&&&&&\\
  $\theta$&0.1128&	0.0128(0.0014)&	0.1079&	0.0079(0.0007)	&0.1080	&0.0080(0.0002)	&0.1027&	0.0027 (0.0001)\\
  $\alpha$&0.7109&	0.4609(0.2184)&	0.7034&	0.4534(0.2088)&	0.6957&	0.4457(0.2000)&	0.6804&	0.4304 (0.1888)\\
  $\delta$&3.0685&	0.5685(1.0447)&	3.0085&	0.5085(0.7195)&	2.9490&	0.4490(0.3138)&	2.8473&	0.3473 (0.1865)\\
\hline
$\alpha=0.75$&&&&&&&&\\
$\theta$&0.1351	&0.0351(0.0065)	&0.1217&	0.0217(0.0017)&	0.1203&	0.0203(0.0023)&	0.1156&	0.0156(0.0004)\\
$\alpha$&0.8487	&0.0987(0.0167)	&0.8414&	0.0914(0.0135)&	0.8318&	0.0818(0.0095)&	0.8190&	0.0690(0.0081)\\
$\delta$&3.1905	&0.6905(2.0102)	&3.1765&	0.6765(1.4733)&	3.0104	&0.5104(0.7818)&	2.8781&	0.3781(0.3592)\\
\hline
\end{tabular}
\end{small}
\end{table}
\newpage
Table 5. Performance of MLE method for PMQLD$(\theta=0.30,\alpha,\delta=2.50)$
\begin{table}[H]
\centering
\begin{small}
  \begin{tabular}{ c|cccccccc} 
  \hline
  &\multicolumn{2}{c}{$n=60$}&\multicolumn{2}{c}{$n=100$}&\multicolumn{2}{c}{$n=200$}&\multicolumn{2}{c}{$n=300$}\\
  \hline
  &MLE&Bias(MSE)&MLE&Bias(MSE)&MLE&Bias(MSE)&MLE&Bias(MSE)\\
  \hline
  $\alpha=0.25$&&&&&&&&\\
  $\theta$&0.3998&	0.0998(0.0329)&	0.3805&	0.0805(0.0240)&	0.3482&	0.0482(0.0060)&	0.3336&	0.0336(0.0030)\\
  $\alpha$&0.7823&	0.5323(0.2928)&	0.7795&	0.5295(0.2873)&	0.7611&	0.5111(0.2636)&	0.7437&	0.4937(0.2508)\\
  $\delta$&3.5545&	1.0545(2.7164)	&3.4774&	0.9774(2.2678)	&3.1584	&0.6584(0.7994)	&3.0583&	0.5583(0.5396)\\
\hline
$\alpha=0.75$&&&&&&&&\\
$\theta$&0.3773&	0.0773(0.0621)&	0.3566&	0.0566(0.0506)&	0.3275&	0.0275(0.0267)&	0.3097&	0.0097(0.0225)\\
$\alpha$&0.8962	&0.1462(0.0280)&	0.8889&	0.1389(0.0221)&	0.8693&	0.1193(0.0186)&	0.8475&	0.0975(0.0175)\\
$\delta$&2.9598	&0.4598(4.7991)	&2.8278&	0.3278(4.4320)&	2.6089&0.1089(2.7618)&	2.5723&	0.0723(2.4811)\\
\hline
\end{tabular}
\end{small}
\end{table}
\subsection{Real-world applications}
In this subsection, we discuss the real-world applications of the proposed mixed Poisson distribution. Three data sets are considered to illustrate whether the proposed distribution is well fitted compared to some other existing competing Poisson mixtures. The best-fitted distribution was selected based on the negative log-likelihood $(-2logL)$, Akaike Information Criterion (AIC), and chi-square goodness of fit statistic. The unknown parameters of the models are estimated by using the MLE method. Tables 6 to 8 summarize all these statistical measures for each data set, and the standard errors of the parameter estimates are reported in parentheses. \\\\
The first data set contains the epileptic seizure counts (Chakraborty, 2010). The sample index dispersion $(\tau)$ of this data set is 1.867. Since $\tau$ value is greater than one, the distribution of this data set is clearly over-dispersed. Also, the skewness and excess kurtosis for this example are 1.239 and 1.680, respectively, which show that the distribution of the data set is positively skewed and leptokurtic. This data set was used to fit the PMQLD, ZMPMQLD, GD, NBD, PLD, GPLD, PQLD, and PGLD. Table 6 presents the estimates of the parameters of distributions and the goodness of fit test. Of all eight distributions, the PMQLD performs well based on the smallest AIC value of 1191.83 and the smallest chi-square value ($\chi^2$) of 2.93 (p-value=0.71).\\\\
The second data set represents the number of roots produced by 270 micro-propagated shoots of the columnar apple cultivar Trajan (Ridout et al., 1998). This is a bimodal data set for which the sample index dispersion, skewness, and excess kurtosis are 3.077, 0.182, and -1.056, respectively. These values indicate that the distribution of the data set is extremely over-dispersed, mild positively skewed, and platykurtic. This data set was also used to fit the same distributions that we used for the first example. Table 7 summarizes the results of parameter estimations and the goodness of fit test. The results show that the PMQLD having AIC=1350.20, $\chi^2=11.75$, p-value=0.47 outperforms clearly than other distributions.\\\\
The third example is used to show the performance of the ZMPMQLD. This data set presents the number of units of consumers' goods purchased by households over 26 weeks (Lindsey, 1995). The proportion of the zeros in the data set is 80.60$\%$, which indicates that there exists inflation of zeros. The sample dispersion index of 4.761 shows that extreme over-dispersion is present. Further, the skewness and excess kurtosis of the data are 3.895 and 16.306, respectively. These results imply that the distribution of the data set is highly positively skewed having a very long tail. The NBD, zero-modified negative binomial distribution (ZMNBD), PLD, zero-modified Poisson-Lindley distribution (ZMPLD), PMQLD, and ZMPMQLD are fitted to the data set. Parameter estimates, chi-square statistics, and p-value of the chi-square goodness of fit test are summarized in Table 8. Based on the results, the ZMPMQLD having AIC=3419.95, $\chi^2=12.22$, p-value=0.06 gives a better fit than the other distributions. Further, when we compare the PMQLD and ZMPMQLD, the likelihood ratio (LR) test statistic for the hypothesis testing $H_{0}:\phi=0$ vs $H_{a}:\phi\neq 0$ for this data set is 50.50, and it is greater than $\chi_{1,0.05}^{2}=3.84$. Then, the parameter estimate $\hat{\phi}$ is significantly different from zero.\\\\ Figure 7 illustrates how the expected values of the proposed distribution adhere with the observed value for the data sets. We can see that the observed values of the first and second data sets are very close to the expected values of the PMQLD, and the observed values of the third data set are very close to the expected values of the ZMPQLD.
\begin{figure}[H]\label{fig:my_label}
        \includegraphics[width=16.0cm, height=09.5cm]{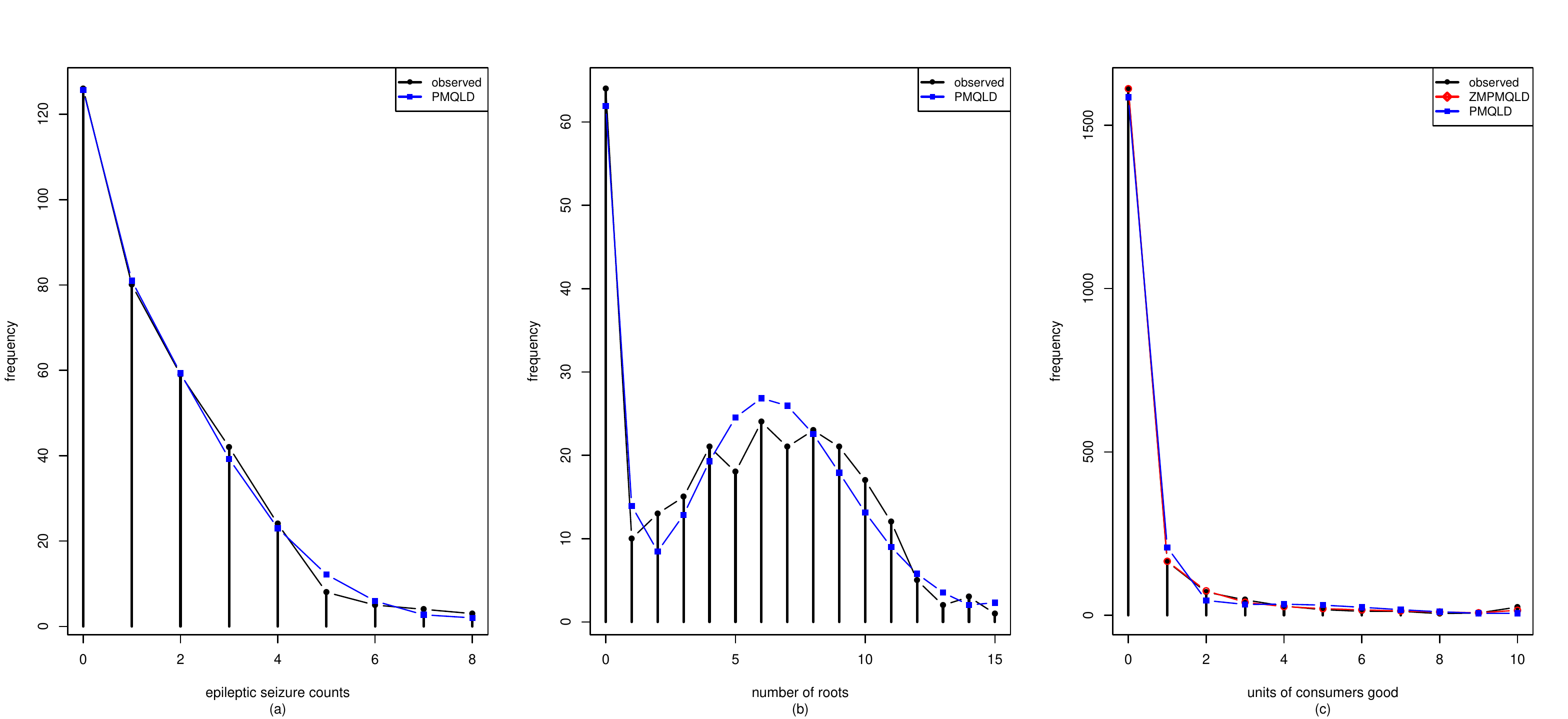}
   \caption{Performance of PMQLD and ZMPMQLD to real-data sets}
   \end{figure}
   	Table 6. Epileptic seizure counts 
   \begin{table}[H]
   \centering
\begin{footnotesize}
  \begin{tabular}{ cccccccccc}
  \hline
 Counts &Observed&\multicolumn{8}{c}{Expected}\\
 &&GD&NBD&PLD&GPLD&PQLD&PGLD&PMQLD&ZMPMQLD\\
  \hline
  0	&126&	137.95&	120.22&	128.72&	121.93&	121.82&	122.85&	125.65&	126.04\\
  1	&80	&83.73	&93.00&	87.14&	90.92&	90.95&	90.08&	80.98&	80.12\\
  2	&59	&50.82&	59.18&	55.26&	58.72&	58.76&	57.85&	59.30&	59.16\\
  3	&42	&30.85&	34.94&	33.63&	35.20&	35.23&	35.20&	39.25&	39.72\\
  4	&24	&18.72&	19.84&	19.89&	20.16&	20.17&	20.53&	22.99&	23.35\\
  5	&8	&11.37&	10.99&	11.52&	11.20&	11.20&	11.54&	12.17&	12.26\\
  6	&5	&6.90&	5.98&	6.57&	6.08&	6.08&	6.27&	5.94&	5.89\\
  7	&4&	4.19&	3.22&	3.70&	3.25&	3.25&	3.31&	2.72&	2.63\\
  8	&3&	6.47&	3.63&	4.57&	3.54&	3.54&	3.37&	2.00&	1.83\\
  Total	&351&	351&	351&	351&	351&	351&	351&	351&	351\\
  \hline
&&$\hat{\theta}=0.65$&$\hat{\beta}=1.00$&	$\hat{\theta}=0.97$&$\hat{\theta}=1.11$&	$\hat{\theta}=1.12$&	$\hat{\theta}=1.57$&	$\hat{\theta}=2.70 $&	$\hat{\phi}=-0.02$\\
&&$\hspace*{0.5cm}(0.04)$&$\hspace*{0.5cm}(0.19)$&$\hspace*{0.5cm}(0.05)$&$\hspace*{0.5cm}(0.13)$&$\hspace*{0.5cm}(0.13)$&$\hspace*{0.5cm}(0.66)$&$\hspace*{0.5cm}(1.26)$&$\hspace*{0.5cm}(0.49)$\\
&&&$\hat{\alpha}=1.55$&&$\hat{\alpha}=2.76$&$\hat{\alpha}=0.38$&	$\hat{\alpha}=1.49$&	$\hat{\alpha}=0.82$&	$\hat{\theta}=3.10$\\
MLE&&&$\hspace*{0.5cm}(0.28)$&&$\hspace*{0.5cm}(2.76)$&$\hspace*{0.5cm}(0.33)$&$\hspace*{0.5cm}(0.57)$&$\hspace*{0.5cm}(0.07)$&$\hspace*{0.5cm}(3.77)$\\
&&&&&&&$\hat{\beta}=3.89$&	$\hat{\delta}=5.89$&	$\hat{\alpha}=0.84$\\
&&&&&&&$\hspace*{0.5cm}(2.38)$&$\hspace*{0.5cm}(2.83)$&$\hspace*{0.5cm}(0.56)$\\
&&&&&&&&&$\hat{\delta}=6.86$\\
&&&&&&&&&$\hspace*{0.5cm}(9.65)$\\
\hline
$\chi^2$&&11.42&	5.67&	5.84&	4.85&	4.86&	4.66&	2.93&	3.22\\
p-value&&0.12&	0.46&	0.56&	0.56&	0.56&	0.46&	0.71&	0.52\\
$-2logL$&&1196.79&	1189.88&	1190.36&	1188.96&	1188.96&	1188.54&	1185.83&	1185.91\\
AIC&&1198.79&	1193.88&	1192.36&	1192.96&	1192.96&	1194.54&	1191.83&	1193.91\\
\hline
\end{tabular}
\end{footnotesize}
\end{table}
Table 7. Number of roots 
\begin{table}[H]
\centering
\begin{footnotesize}
  \begin{tabular}{ cccccccccc}
  \hline
 Counts &Observed&\multicolumn{8}{c}{Expected}\\
 &&GD&NBD&PLD&GPLD&PQLD&PGLD&PMQLD&ZMPMQLD\\
  \hline
  0	&64	&44.62&	36.87&	31.09&	35.46&	35.45&	82.81&	61.93&	64.77\\
  1	&10&	37.25	&36.05&	32.94&	34.00&	33.99&	17.81&	13.92&	9.24\\
  2	&13	&31.09	&32.16&	31.79&	31.19&	31.19&	15.31&	8.47&	13.45\\
  3&	15&	25.95&	27.77&	29.06&	27.78&	27.77&	16.46&	12.85&	18.44\\
4&	21	&21.66&	23.58&	25.63&	24.20&	24.20&	17.53&	19.30&	22.23\\
5&	18	&18.08	&19.83&	22.04&	20.74&	20.74&	17.70&	24.53&	23.89\\
6&	24&	15.09&	16.56&	18.60&	17.55&	17.55&	16.95&	26.85&	23.39\\
7&	21&	12.60&	13.76&	15.48&	14.69&	14.69&	15.53&	25.94&	21.27\\
8&	23&	10.52&	11.39&	12.73&	12.20&	12.20&	13.70&	22.55&	18.20\\
9&	21&	8.78&	9.40&	10.37&	10.05&	10.05&	11.72&	17.90&	14.81\\
10&	17&	7.33&	7.74&	8.39&	8.23&	8.24&	9.76&	13.13&	11.56\\
11	&12&	6.12&	6.37&	6.74&	6.71&	6.71&	7.96&	8.99&	8.70\\
12	&5&	5.11&	5.23&	5.38&	5.44&	5.45&	6.36&	5.78&	6.35\\
13&	2&	4.26&	4.28&	4.27&	4.40&	4.40&	5.01&	3.52&	4.51\\
14&	3&	3.56&	3.51&	3.38&	3.54&	3.54&	3.88&	2.03&	3.13\\
$\geq 15$&1&	17.98&	15.50&	12.11&	13.82&	13.83&	11.51&	2.31&	6.06\\
\hline
Total&	270&	270&	270	&270	&270&	270&	270&	270&	270\\
\hline
&&$\hat{\theta}=0.12$&$\hat{\beta}=0.24$&	$\hat{\theta}=0.35$&$\hat{\theta}=0.37$&	$\hat{\theta}=0.32$&	$\hat{\theta}=0.59$&	$\hat{\theta}=4.23 $&	$\hat{\phi}=-0.21$\\
&&$\hspace*{0.5cm}(0.01)$&$\hspace*{0.5cm}(0.03)$&$\hspace*{0.5cm}(0.02)$&$\hspace*{0.5cm}(0.03)$&$\hspace*{0.5cm}(0.03)$&$\hspace*{0.5cm}(0.08)$&$\hspace*{0.5cm}(1.36)$&$\hspace*{0.5cm}(0.04)$\\
&&&$\hat{\alpha}=1.21$&&$\hat{\alpha}=0.47$&$\hat{\alpha}=0.68$&	$\hat{\alpha}=0.22$&	$\hat{\alpha}=0.73$&	$\hat{\theta}=1.10$\\
MLE&&&$\hspace*{0.5cm}(0.16)$&&$\hspace*{0.5cm}(0.26)$&$\hspace*{0.5cm}(0.27)$&$\hspace*{0.5cm}(0.12)$&$\hspace*{0.5cm}(0.04)$&$\hspace*{0.5cm}(0.24)$\\
&&&&&&&$\hat{\beta}=4.25$&	$\hat{\delta}=29.34$&	$\hat{\alpha}=0.40$\\
&&&&&&&$\hspace*{0.5cm}(0.66)$&$\hspace*{0.5cm}(9.28)$&$\hspace*{0.5cm}(0.15)$\\
&&&&&&&&&$\hat{\delta}=7.38$\\
&&&&&&&&&$\hspace*{0.5cm}(1.71)$\\
\hline
$\chi^2$&&121.92&	120.76&	117.44&	110.58&	110.58&	46.72&	11.74&	15.85\\
p-value&&0.00&	0.00&	0.00&	0.00&	0.00&	0.00&	0.466&	0.147\\
$-2logL$&&1464.90&	1462.63&	1454.10&	1451.45&	1451.45&	1384.21	&1344.20&	1350.92\\
AIC&&1466.90&	1466.63&	1456.10&	1455.45&	1455.45&	1390.21&	1350.20&	1358.92\\
\hline
\end{tabular}
\end{footnotesize}
\end{table}
Table 8. Units of consumers good 
\begin{table}[H]
\centering
\begin{footnotesize}
  \begin{tabular}{ ccccccccc}
  \hline
 Counts &Observed&\multicolumn{6}{c}{Expected}\\
 &&NBD&	ZMNBD&	PLD	&ZMPLD&	PMQLD&	ZMPMQLD\\
  \hline
  0	&1612&	1239.62&	1617.92&	1272.56	&1611.94&	1585.72&	1611.78\\
  1	&164&	240.36&	146.26&	470.84&	122.00&	207.56&	165.09\\
  2	&71	&130.05&	80.89&	168.06&	88.57&	45.30&	74.66\\
  3	&47&	85.41&	50.02&	58.50&	61.44&	32.99&	39.88\\
  4	&28&	61.04&	32.56&	19.99&	41.33&	33.90&	26.36\\
  5	&17	&45.74&	21.83&	6.73&	27.20&	30.93&	20.24\\
  6	&12	&35.33&	14.93&	2.24&	17.59&	24.36&	16.34\\
  7	&12	&27.88	&10.34&	0.74&	11.23&	16.90&	13.03\\
  8	&5&	22.35&	7.24&	0.24&	7.10&	10.53&	10.03\\
  9&	7&	18.12&	5.11&	0.07&	4.45&	5.97&	7.41\\
  10&	25&	94.10&	12.90&	0.03&	7.15&	5.84&	15.18\\
\hline
Toatal&	2000&	2000&	2000&	2000&	2000&	2000&	2000\\
\hline
&&$\hat{\beta}=0.21$&$\hat{\phi}=0.33$&	$\hat{\theta}=2.33$&$\hat{\phi}=0.73$&	$\hat{\theta}=7.00$&$\hat{\phi}=-0.61$\\
&&$\hspace*{0.5cm}(0.21)$&$\hspace*{0.5cm}(0.30)$&$\hspace*{0.5cm}(0.07)$&$\hspace*{0.5cm}(0.01)$&$\hspace*{0.5cm}(0.67)$&$\hspace*{0.5cm}(0.03)$\\
&&$\hat{\alpha}=0.12$&$\hat{\beta}=0.23$&&$\hat{\theta}=0.75$&$\hat{\alpha}=2.12$&$\hat{\theta}=1.45$\\
MLE&&$\hspace*{0.5cm}(0.01)$&$\hspace*{0.5cm}(0.04)$&&$\hspace*{0.5cm}(0.04)$&$\hspace*{0.5cm}(0.07)$&$\hspace*{0.5cm}(0.25)$\\
&&&$\hat{\alpha}=0.20$&&&$\hat{\delta}=32.88$&$\hat{\alpha}=1.78$\\
&&&$\hspace*{0.5cm}(0.13)$&&&$\hspace*{0.5cm}(2.81)$&$\hspace*{0.5cm}(0.11)$\\
&&&&&&&$\hat{\delta}=8.51$\\
&&&&&&&$\hspace*{0.5cm}(1.32)$\\
\hline
$\chi^2$&&311.64&	18.86&	17992.93&	77.89&	110.99&	12.22\\
p-value&&0.00&	0.01&	0.00&	0.00&	0.00&	0.06\\
$-2logL$&&3800.90&	3427.16&	4216.21&	3455.33&	3462.45&	3411.95\\
AIC&&3804.90	&3433.16&	4218.21&	3459.33&	3468.45&	3419.95\\
\end{tabular}
\end{footnotesize}
\end{table}
\section{Conclusion}
This paper proposes an alternative mixed Poisson distribution with its zero-modified version to model the over-dispersed count data. Explicit expressions of the pmf, hazard rate function, moments, mean, variance, skewness, and kurtosis were derived for the proposed distribution. Its pmf possesses to be either unimodal or bimodal, and hazard rate function presents monotonic increasing, decreasing, and bathtub shapes. The kurtosis and the variance-to-mean ratio functions of the new distribution indicate that the distribution can capture various ranges of right tail weights as well as the index of dispersions. Further, its structural properties show that the new distribution is much more flexible than its predecessors, negative binomial, geometric, and Poisson-Lindley distributions. The maximum likelihood method was employed to estimate the parameters of the distribution, and the observed information matrix has also been derived. The proposed distribution and some other competing Poisson mixtures have been fitted to three real-world data sets. The results show that the proposed distribution could provide a better fit than a set of common Poisson mixtures considered in these applications.  

\everypar = {
\parindent=0pt
\hangindent=8mm
\hangafter=1
}\noindent 

Albrecht, P. (1984). Laplace Transforms, Mellin Transforms and Mixed Poisson Processes. \textit{Scandinavian Actuarial Journal}, 11(1), 58-64. https://doi.org/10.1080/03461238.1984.10413753

Bhati, D., Sastry, D.V.S., $\&$ Qadri, P.Z. (2015). A New Generalized Poisson Lindley Distribution, applications and Properties. \textit{Austrian Journal of Statistics}, 44(4), 35-51. https://doi.org/10.17713/ajs.v44i4.54

Bhattacharya, S.K. (1967). A Result in Accident Proneness. \textit{Biometrika}, 54(1), 324–325.\\ https://doi.org/10.1093/biomet/54.1-2.324

Cameron, A.C., $\&$ Trivedi, P.K. (2013). \textit{Regression Analysis of Count Data}. (2nd Ed.) Cambridge University Press, New York, USA. https://doi.org/10.1017/CBO9781139013567

Chakraborty, S. (2010). On Some Distributional Properties of the Family of Weighted Generalized Poisson Distribution.  \textit{Communications in Statistics - Theory and Methods},
39(15), 2767-2788.\\ https://doi.org/10.1080/03610920903129141

Elbatal, I., Merovci, F., $\&$ Elgarhy, M. (2013). A new generalized Lindley distribution, \textit{Mathematical Theory and Modeling}, 3(13), 30-47.

Feller, W. (1943). On a Generalized Class of Contagious Distributions. \textit{ Annals of Mathematical Statistics}, 14(12), 389–400. https://www.jstor.org/stable/2235926

Ghitany, M.E, Atieh, B., $\&$ Nadarajah, S. (2008). Zero-truncated Poisson- Lindley Distribution
and Its Application. \textit{Mathematics and Computers in Simulation}, 79(3), 279-287.\\ https://doi.org/10.1016/j.matcom.2007.11.021

Greenwood, M., $\&$ Yule, G.U. (1920). An inquiry into the
nature of frequency distributions representative of
multiple happenings with particular reference to the
occurrence of multiple attacks of disease or of
repeated accidents, \textit{Journal of the Royal Statistical Society}, 83(2), 255-279. \\https://www.jstor.org/stable/2341080

Grine, R., $\&$ Zeghdoudi, H. (2017). On Poisson quasi-Lindley distribution and its applications. \textit{ J. of Modern Applied Statistical Methods}, 16(2), 403-417.

Irwin, J. (1975). The Generalized Waring Distribution Parts I, II, III. \textit{Journal of the Royal Statistical Society A}, 138(1), 18–31
(Part I), 204–227 (Part II), 374–384 (Part III). 

Johnson, N.L., Kotz, S., $\&$ Kemp, A.W. (1992). \textit{Univariate Discrete Distributions}. (2nd Ed.) John Wiley and Sons, New York, USA.

Lindley, D.V. (1958). Fiducial Distributions and Bayes' Theorem. \textit{Journal of the Royal Statistical Society}, 20(1), 102-107. https://www.jstor.org/stable/2983909

Lindsey, J.K. (1995). \textit{Modelling Frequency and Count Data}. (1st Ed.) Clarendon Press, UK.

Lynch, J. (1988). Mixtures, Generalized Convexity and Balayages, \textit{ Scandinavian Journal of Statistics}, 15(3), 203-210. 

McCullagh, P., $\&$ Nelder, J. (1989). \textit{Generalized Linear Models}. (2nd Ed.) Chapman and Hall, New York, USA.

Ong, S.H., $\&$ Muthaloo, S. (1995). A Class of Discrete Distributions Suited to Fitting Very Long Tailed Data. \textit{ Communication in Statistics-Simulation and Computation}, 24(4), 929–945.\\ https://doi.org/10.1080/03610919508813285

Ridout, M., Demktrio, C.G.B, $\&$ Hinde, J. (1998). Models for count data with many zeros. Invited paper presented at the Nineteenth International Biometric Conference,
Capetown, South Africa. 

Ruohonen, M. (1988). A Model for the Claim Number Process. \textit{ASTIN Bulletin}, 18(1), 57–68.\\ https://doi.org/10.2143/AST.18.1.2014960

Sankaran, M. (1970). The discrete Poisson-Lindley distribution. \textit{Biometrics}, 26(1), 145-149.\\ https://www.jstor.org/stable/2529053

Shaked, M. (1980). On Mixtures from Exponential Families. \textit{ Journal of the Royal Statistical Society B}, 42(2), 192–198.  https://doi.org/10.1111/j.2517-6161.1980.tb01118.x

Shanker, R. $\&$ Mishra, A. (2013a). A Quasi Lindley Distribution. \textit{ African Journal of Mathematics and Computer Science Research}, 6(4), 64-71.

Shanker, R., Sharma, S., $\&$ Shanker, R. (2013b). A Two-Parameter Lindley Distribution for Modeling Waiting and Survival Times Data. \textit{Applied Mathematics}, 4(2), 363-368. 

Tharshan,  R. $\&$ Wijekoon,  P. (2020a). Location Based Generalized Akash Distribution:  Properties and Applications, \textit{Open Journal of Statistics}, 10(2), 163-187. 

Tharshan,  R. $\&$ Wijekoon,  P. (2020b). A comparison study on a new five-parameter generalized Lindley distribution with its sub-models. \textit{Statistics in Transition new Series}, 21(2), 89-117.\\ https://doi.org/10.21307/stattrans-2020-015

Tharshan,  R. and Wijekoon,  P. (2021). A modification of the Quasi Lindley distribution. \textit{Open Journal of Statistics}, 11(3), 369-392. 

Willmot, G.E. (1993). On Recursive Evaluation of Mixed Poisson Probabilities and Related Quantities, \textit{Scandinavian Actuarial Journal}, 18(2), 114–133.

Wongrin, W. and Bodhisuwan, W. (2016). The Poisson-generalised Lindley distribution and its applications. \textit{Songklankarin J.Sci. Technol}, 38(6), 645-66.

Xavier, D., Santos-Neto, M., Bourguignon, M., $\&$ Tomazella, V. (2018). Zero-Modified Poisson-Lindley
distribution with applications in zero-inflated and zero-deflated count data. \textit{arXiv preprint}. arXiv:1712.04088.\vspace*{2cm}
\begin{itemize}
\item [] \textbf{Appendix A}: Proof of proposition 2:\\\\
\textbf{Proof.} 
\\\\
From equation (9):\\\\
\begin{center}$f(\lambda|x)=\displaystyle\frac{\displaystyle\Pi_{i=1}^{n}\displaystyle\frac{e^{-\lambda}\lambda^{x_{i}}}{x_{i}!}\bigg(\displaystyle\frac{\alpha^3\theta e^{-\theta\lambda}}{\alpha^3+1}+\displaystyle\frac{\theta^\delta\lambda^{\delta-1} e^{-\theta\lambda}}{(\alpha^3+1)\Gamma(\delta)}\bigg)}{\displaystyle\int_{0}^{\infty}\displaystyle\Pi_{i=1}^{n}\displaystyle\frac{e^{-\lambda}\lambda^{x_{i}}}{x_{i}!}\bigg(\displaystyle\frac{\alpha^3\theta e^{-\theta\lambda}}{\alpha^3+1}+\displaystyle\frac{\theta^\delta\lambda^{\delta-1} e^{-\theta\lambda}}{(\alpha^3+1)\Gamma(\delta)}\bigg)d\lambda}$\end{center}\begin{center}$=\displaystyle\frac{\Gamma(\delta)\alpha^3 e^{-\lambda(n+\theta)}\lambda^{(\sum_{i=1}^{n}x_{i})}(n+\theta)^{(\sum_{i=1}^{n}x_{i}+\delta)}+\theta^{\delta-1}(n+\theta)^{(\sum_{i=1}^{n}x_{i}+\delta)}\lambda^{(\sum_{i=1}^{n}x_{i}+\delta-1)}e^{-\lambda(n+\theta)}}{\alpha^3\Gamma(\delta)(n+\theta)^{(\delta-1)}\Gamma(\sum_{i=1}^{n}x_{i}+1)+\theta^{\delta-1}\Gamma(\sum_{i=1}^{n}x_{i}+\delta)}$\end{center}
\begin{center}$=\displaystyle\frac{\alpha^3\Gamma(\delta)(n+\theta)^{(\delta-1)}\Gamma(\sum_{i=1}^{n}x_{i}+1)}{\alpha^3\Gamma(\delta)(n+\theta)^{(\delta-1)}\Gamma(\sum_{i=1}^{n}x_{i}+1)+\theta^{\delta-1}\Gamma(\sum_{i=1}^{n}x_{i}+\delta)}$gamma$(\sum_{i=1}^{n}x_{i}+1,n+\theta)$\end{center}\begin{center}
$+\displaystyle\frac{\theta^{\delta-1}\Gamma(\sum_{i=1}^{n}x_{i}+\delta)}{\alpha^3\Gamma(\delta)(n+\theta)^{(\delta-1)}\Gamma(\sum_{i=1}^{n}x_{i}+1)+\theta^{\delta-1}\Gamma(\sum_{i=1}^{n}x_{i}+\delta)}$gamma$(\sum_{i=1}^{n}x_{i}+\delta,n+\theta).$\end{center}
\item [] \textbf{Appendix B}: Elements of the observed information matrix, $I(\theta,\alpha,\delta)$ defined in subsection 5.2.1:\\\\
Let us define $T_{1}, T_{2}, T_{3}, T_{4}, T_{5},T_{6},T_{7},T_{8},T_{9},T_{10},T_{11},T_{12},T_{13}$ and $T_{14}$ as follows
\begin{center}$T_{1}=\Gamma(\delta)\Gamma(x_{i}+1)\alpha^3(\delta-1)(1+\theta)^{\delta-2}+\Gamma(x_{i}+\delta)(\delta-1)\theta^{\delta-2},$\end{center}
\begin{center}$T_{2}=\Gamma(\delta)\Gamma(x_{i}+1)\alpha^3(1+\theta)^{\delta-1}+\theta^{\delta-1}\Gamma(x_{i}+\delta),$\end{center}
\begin{center}$T_{3}=3\alpha^2\Gamma(\delta)\Gamma(x_{i}+1)(1+\theta)^{\delta-1},$\end{center}
\begin{center}$T_{4}=\Gamma(x_{i}+1)\alpha^3(\Gamma(\delta)(1+\theta)^{\delta-1}log(1+\theta)+(1+\theta)^{\delta-1}\Gamma(\delta)\psi(\delta))+\Gamma(x_{i}+\delta)\theta^{\delta-1}(\log(\theta)+\psi(x_{i}+\delta)),$ \end{center}
\begin{center}$T_{5}=\Gamma(\delta)\Gamma(x_{i}+1)\alpha^3(1+\theta)^{\delta-3}+\Gamma(x_{i}+\delta)\theta^{\delta-3},$\end{center}
\begin{center}$T_{6}=\Gamma(\delta)\Gamma(x_{i}+1)\alpha^3(1+\theta)^{\delta-2}+\Gamma(x_{i}+\delta)\theta^{\delta-2},$\end{center}
\begin{center}$T_{7}=log(1+\theta)(\Gamma(\delta)(1+\theta)^{\delta-1}log(1+\theta)+(1+\theta)^{\delta-1}\Gamma(\delta)\psi(\delta))$\end{center}
\begin{center}$T_{8}=(1+\theta)^{\delta-1}(\Gamma(\delta)\psi_{1}(\delta)+(\psi(\delta))^2\Gamma(\delta))+\Gamma(\delta)\psi(\delta)(1+\theta)^{\delta-1}log(1+\theta),$\end{center}
\begin{center}$T_{9}=\Gamma(x_{i}+\delta)\theta^{\delta-1}log(\theta)+\theta^{\delta-1}\Gamma(x_{i}+\delta)\psi(x_{i}+\delta),$\end{center}
\begin{center}$T_{10}=\Gamma(x_{i}+\delta)\theta^{\delta-1}\psi_{1}(x_{i}+\delta),$\end{center}
\begin{center}$T_{11}=\Gamma(x_{i}+1)\alpha^3(\Gamma(\delta)(1+\theta)^{\delta-1}log(1+\theta)+(1+\theta)^{\delta-1}\Gamma(\delta)\psi(\delta)),$\end{center}
\begin{center}$T_{12}=\theta^{\delta-1}\Gamma(x_{i}+\delta)\psi(x_{i}+\delta)+\Gamma(x_{i}+\delta)\theta^{\delta-1}log(\theta),$\end{center}
\begin{center}$T_{13}=\Gamma(\delta)((1+\theta)^{\delta-2}+log(1+\theta)(\delta-1)(1+\theta)^{\delta-2})+(\delta-1)(1+\theta)^{\delta-2}\Gamma(\delta)\psi(\delta),$\end{center} and
\begin{center}$T_{14}=(log(\theta)+\psi(x_{i}+\delta))\Gamma(x_{i}+\delta)(\delta-1)\theta^{\delta-2}.$\end{center}
Then, the second order partial derivatives of the log-likelihood function are as follows\begin{center}
$\displaystyle\frac{\partial^2l(\theta,\alpha,\delta|x)}{\partial\theta^2}=\displaystyle\frac{-n}{\theta^2}+\sum_{i=1}^{n}\frac{x_{i}+\delta}{(1+\theta)^2}+\displaystyle\sum_{i=1}^{n}\frac{T_{2}(\delta-1)(\delta-2)T_{5}-T_{1}(\delta-1)T_{6}}{T_{2}^2},$\end{center}
\begin{center}$\displaystyle\frac{\partial^2l(\theta,\alpha,\delta|x)}{\partial\alpha^2}=\displaystyle\sum_{i=1}^{n}\frac{T_{2}(6\alpha\Gamma(\delta)\Gamma(x_{i}+1)(1+\theta)^{\delta-1})-T_{3}(3\alpha^2\Gamma(\delta)\Gamma(x_{i}+1)(1+\theta)^{\delta-1})}{T_{2}^{2}}-\displaystyle\frac{3n\alpha(2(\alpha^2+1)-3\alpha^3)}{(\alpha^3+1)^2},$\end{center}
\begin{center}$\displaystyle\frac{\partial^2l(\theta,\alpha,\delta|x)}{\partial\delta^2}=\displaystyle\sum_{i=1}^{n}\displaystyle\frac{T_{2}(\Gamma(x_{i}+1)\alpha^3(T_{7}+T_{8})+(log(\theta)+\psi(x_{i}+\delta))T_{9}+T_{10})-T_{4}(T_{11}+T_{12})}{T_{2}^2}-n\psi_{1}(\delta),$\end{center}
\begin{center}$\displaystyle\frac{\partial^2l(\theta,\alpha,\delta|x)}{\partial\theta\partial\alpha}=\displaystyle\sum_{i=1}^{n}\frac{T_{2}(3\alpha^2\Gamma(\delta)\Gamma(x_{i}+1)(\delta-1)(1+\theta)^{\delta-1})-T_{1}T_{3}}{T_{1}^2},$\end{center}
\begin{center}$\displaystyle\frac{\partial^2l(\theta,\alpha,\delta|x)}{\partial\delta\partial\alpha}=\displaystyle\sum_{i=1}^{n}\frac{T_{2}(3\alpha^2\Gamma(x_{i}+1)(\Gamma(\delta)(1+\theta)^{\delta-1}log(1+\theta)+(1+\theta)^{\delta-1}\Gamma(\delta)\psi(\delta)))-T_{4}T_{3}}{T_{2}^2},$\end{center} and \begin{center}
$\displaystyle\frac{\partial^2l(\theta,\alpha,\delta|x)}{\partial\delta\partial\theta}=\displaystyle\sum_{i=1}^{n}\frac{T_{2}(\Gamma(x_{i}+1)\alpha^3T_{13}+\Gamma(x_{i}+\delta)\theta^{\delta-2}+T_{14})-T_{4}T_{1}}{T_{2}^2}-\displaystyle\frac{n}{1+\theta},$\end{center}
where $\psi_{1}(s)$ is the trigamma function and defined as $\psi_{1}(s)=\displaystyle\frac{d^2}{ds^2}log(\Gamma(s))=\displaystyle\sum_{i=1}^{n}\frac{1}{(s+k)^2}.$
\item []\textbf{Appendix C}: Partial derivatives of the log-likelihood function of equation (16), $l(\phi,\theta,\alpha,\delta|x)$ (score functions):
\begin{center}$\displaystyle\frac{\partial l(\phi,\theta,\alpha,\delta|x)}{\partial\phi}=\displaystyle\frac{n_{0}((\alpha^3+1)(1+\theta)^\delta-\theta((1+\theta)^{\delta-1}\alpha^3+\theta^{\delta-1}))}{\phi(\alpha^3+1)(1+\theta)^\delta+(1-\phi)\theta((1+\theta)^{\delta-1}\alpha^3+\theta^{\delta-1})}-\displaystyle\frac{n-n_{0}}{1-\phi},$\end{center}
\begin{center}$\displaystyle\frac{\partial l(\phi,\theta,\alpha,\delta|x)}{\partial\theta}=\displaystyle\frac{n-n_{0}}{\theta}+\sum_{i=1}^{n}(1-I_{(X=0)}(x_{i}))\frac{\Gamma(\delta)\Gamma(x_{i}+1)\alpha^3(\delta-1)(1+\theta)^{\delta-2}+(\delta-1)\theta^{\delta-2}\Gamma(x_{i}+\delta)}{\Gamma(\delta)\Gamma(x_{i}+1)\alpha^3(1+\theta)^{\delta-1}+\theta^{\delta-1}\Gamma(x_{i}+\delta)}+$\end{center}\begin{center}$\displaystyle\frac{n_{0}(1-\phi)\bigg((1+\theta)^\delta(\alpha^3(\theta(\delta-1)(1+\theta)^{\delta-2}+(1+\theta)^{\delta-1})+\delta\theta^{\delta-1})-\theta((1+\theta)^{\delta-1}\alpha^3+\theta^{\delta-1})\delta(1+\theta)^{\delta-1}\bigg)}{(1+\theta)^\delta(\phi(\alpha^3+1)(1+\theta)^\delta+(1-\phi)\theta((1+\theta)^{\delta-1}\alpha^3+\theta^{\delta-1}))}$\end{center}\begin{center}$-\sum_{i=1}^{n}(1-I_{(X=0)}(x_{i}))\displaystyle\frac{(x_{i}+\delta)}{1+\theta},$\end{center}
\begin{center}$\displaystyle\frac{\partial l(\phi,\theta,\alpha,\delta|x)}{\partial\alpha}=\frac{n_{0}(1-\phi)\theta((\alpha^3+1)(3\alpha^2(1+\theta)^{\delta-1})-3\alpha^2((1+\theta)^{\delta-1}\alpha^3+\theta^{\delta-1}))}{(\phi(\alpha^3+1)(1+\theta)^\delta+(1-\phi)\theta((1+\theta)^{\delta-1}\alpha^3+\theta^{\delta-1}))(\alpha^3+1)}-\displaystyle\frac{3\alpha^2(n-n_{0})}{\alpha^3+1}$\end{center}\begin{center}$+\sum_{i=1}^{n}(1-I_{(X=0)}(x_{i}))\displaystyle\frac{3\alpha^2\Gamma(\delta)\Gamma(x_{i}+1)(1+\theta)^{\delta-1}}{\Gamma(\delta)\Gamma(x_{i}+1)\alpha^3(1+\theta)^{\delta-1}+\theta^{\delta-1}\Gamma(x_{i}+\delta)},$\end{center} and\\\\
\begin{center}$\displaystyle\frac{\partial l(\phi,\theta,\alpha,\delta|x)}{\partial\delta}=-(n-n_{0})(\psi(\delta)+log(1+\theta))+\sum_{i=1}^{n}(1-I_{(X=0)}(x_{i}))\frac{C+D}{\Gamma(\delta)\Gamma(x_{i}+1)\alpha^3(1+\theta)^{\delta-1}+\theta^{\delta-1}\Gamma(x_{i}+\delta)}$\end{center}\begin{center}$+\displaystyle\frac{n_{0}(1-\phi)\theta\bigg((1+\theta)^\delta(\alpha^3(1+\theta)^{\delta-1}log(1+\theta)+\theta^{\delta-1} log(\theta))-((1+\theta)^{\delta-1}\alpha^3+\theta^{\delta-1})(1+\theta)^\delta log(1+\theta)\bigg)}{(\phi(\alpha^3+1)(1+\theta)^\delta+(1-\phi)(\theta(1+\theta)^{\delta-1}\alpha^3+\theta^{\delta-1}))(1+\theta)^{\delta}},$\end{center}
where $C=\Gamma(x_{i}+1)\alpha^3(\Gamma(\delta)(1+\theta)^{\delta-1} log(1+\theta)+(1+\theta)^{\delta-1}\Gamma(\delta)\psi(\delta))$, and \\\\
$D=\Gamma(x_{i}+\delta)\theta^{\delta-1}log(\theta)+\theta^{\delta-1}\Gamma(x_{i}+\delta)\psi(x_{i}+\delta)$.
\end{itemize}
\end{document}